\documentclass[journal,twoside]{IEEEtran}

\usepackage{subfigure}
\usepackage{graphicx}
\usepackage{booktabs}
\usepackage{longtable}
\usepackage{amssymb}
\usepackage{arabtex}
\usepackage{amsmath}
\usepackage{cite}
\usepackage{multirow}
\usepackage{url}
\usepackage{hyperref}

\begin{document}

\title{Convolutional Neural Network-Based Block Up-sampling for Intra Frame Coding}

\author{Yue~Li,
Dong~Liu, \IEEEmembership{Member, IEEE,}
Houqiang~Li, \IEEEmembership{Senior Member, IEEE,}
Li~Li, \IEEEmembership{Member, IEEE,}
Feng~Wu,~\IEEEmembership{Fellow,~IEEE,}
Hong~Zhang,
and Haitao~Yang
\thanks{Date of current version July 13, 2017. This work was supported by the National Program on Key Basic Research Projects (973 Program) under Grant 2015CB351803, by the Natural Science Foundation of China (NSFC) under Grants 61390512, 61325009, 61425026, and 61632001, and by the Fundamental Research Funds for the Central Universities under Grant WK3490000001.

Y. Li, D. Liu, H. Li, and F. Wu are with the CAS Key Laboratory of Technology in Geo-Spatial Information Processing and Application System, University of Science and Technology of China, Hefei 230027, China (e-mail: lytt@mail.ustc.edu.cn; dongeliu@ustc.edu.cn; lihq@ustc.edu.cn; fengwu@ustc.edu.cn).

L. Li is with University of Missouri-Kansas City, 5100 Rockhill Road, Kansas City, MO 64111, USA (e-mail: lil1@umkc.edu).

H. Zhang and H. Yang are with the Media Technology Laboratory, Central Research Institute of Huawei Technologies Co., Ltd, Shenzhen 518129, China (e-mail: summer.zhanghong@huawei.com; haitao.yang@huawei.com).

Copyright \copyright ~2017 IEEE. Personal use of this material is permitted. However, permission to use this material for any other purposes must be obtained from the IEEE by sending an email to pubs-permissions@ieee.org.}
}

\markboth{IEEE Transactions on Circuits and Systems for Video Technology}%
{Li \MakeLowercase{\textit{et al.}}: Convolutional Neural Network-Based Block Up-sampling for Intra Frame Coding}

\maketitle

\begin{abstract}
Inspired by the recent advances of image super-resolution using convolutional neural network (CNN), we propose a CNN-based block up-sampling scheme for intra frame coding. A block can be down-sampled before being compressed by normal intra coding, and then up-sampled to its original resolution. Different from previous studies on down/up-sampling-based coding, the up-sampling methods in our scheme have been designed by training CNN instead of hand-crafted. We explore a new CNN structure for up-sampling, which features deconvolution of feature maps, multi-scale fusion, and residue learning, making the network both compact and efficient. We also design different networks for the up-sampling of luma and chroma components, respectively, where the chroma up-sampling CNN utilizes the luma information to boost its performance. In addition, we design a two-stage up-sampling process, the first stage being within the block-by-block coding loop, and the second stage being performed on the entire frame, so as to refine block boundaries. We also empirically study how to set the coding parameters of down-sampled blocks for pursuing the frame-level rate-distortion optimization. Our proposed scheme is implemented into the High Efficiency Video Coding (HEVC) reference software, and a comprehensive set of experiments have been performed to evaluate our methods. Experimental results show that our scheme achieves significant bits saving compared with HEVC anchor especially at low bit rates, leading to on average 5.5\% BD-rate reduction on common test sequences and on average 9.0\% BD-rate reduction on ultra high definition (UHD) test sequences.
\end{abstract}


\IEEEpeerreviewmaketitle


\section{Introduction}
Video resolution keeps increasing in the past three decades along with the development of new video capture and display devices. The International Telecommunication Union has approved ultra high definition (UHD) television as standard, defining both 4K and 8K that lead to a new level of spatial resolution \cite{sugawara2014ultra}. While UHD video applications, such as home theater, provide users with further enhanced experience and become increasingly popular, they raise even bigger challenges to the video storage and transmission systems. Accordingly, video coding methods have been more and more focused on high definition videos. The state-of-the-art video coding standard, High Efficiency Video Coding (HEVC), supports up to 8K resolution \cite{sullivan2012overview}. However, there is still necessity to further increase the compression efficiency for UHD videos, especially in scenarios where bandwidth is limited for video transmission.

Although the video capture and display devices enable higher resolution, such resolution may not be necessary to carry the important visual information in videos. Thus, it is a well known strategy to down-sample videos prior to encoding and to up-sample the decoded videos for reconstruction \cite{bruckstein2003down,takahashi2011rate,molina2006toward,barreto2007region,shen2011down,wu2009image,lin2006adaptive, nguyen2008adaptive}. Previous studies have shown that using low-resolution version during coding performs better than direct coding of full-resolution videos in low bit rate scenarios \cite{bruckstein2003down,takahashi2011rate}. Moreover, the critical resolution for reconstructing signal is known to be dependent on the spatial frequency of image/video, but different regions of natural images/videos have very different spatial frequency components. Then, several researches have been performed on spatially variant sampling rates for down/up-sampling-based image/video coding \cite{lin2006adaptive, nguyen2008adaptive}.

The up-sampling process plays a key role in down/up-sampling-based video coding as it immediately decides the quality of the final reconstructed videos. Some researches then have been focused on devising more efficient up-sampling methods \cite{molina2006toward,barreto2007region,shen2011down,wu2009image}. Actually, image up-sampling is a classic research topic and has been extensively studied in the literature of image processing, where it is also termed super-resolution (SR). Typical image SR methods can be categorized into interpolation-based, reconstruction-based, and learning-based \cite{yang2010image}, and some of these methods were borrowed into video coding. For example, Shen \emph{et al.} proposed a down/up-sampling-based video coding scheme, where the up-sampling method is a learning-based one that enhances the current low-resolution reconstructed image from the information of an external high-resolution image set \cite{shen2011down}. Nonetheless, most of the previous studies on down/up-sampling-based video coding adopt fixed, hand-crafted interpolation filters rather than many advanced SR methods, partially due to the consideration of computational complexity.

Recently, learning-based image SR using convolutional neural network (CNN) has demonstrated remarkable progress. Dong \emph{et al.} first proposed a CNN-based SR method known as SRCNN, which clearly outperforms the previous rivals in the single image SR task \cite{dong2014learning}. Since then, several CNN-based SR methods have been developed and shown to achieve further performance boost \cite{dong2016accelerating, wang2016end, kim2016accurate, kim2016deeply}.

Inspired by the abovementioned advances, in this paper, we propose a CNN-based block up-sampling scheme for intra frame coding. While it is conceptually natural to replace the hand-crafted interpolation filters with the trained CNN models for better quality, there are lots of issues to investigate when implementing a down/up-sampling-based coding scheme with CNN. First of all, we propose to perform block-level down/up-sampling instead of the entire frame, since different regions have variant local features and then need different sampling rates. Specifically in this work, compliant with the HEVC standard, the basic unit for down/up-sampling is the coding tree unit (CTU). Each CTU can be compressed at its full resolution, or down-sampled by a factor of 2, compressed at low resolution, and then up-sampled. Note that we adopt two different sampling rates here, i.e. $1 \times 1$ and $1/2\times 1/2$, but extension to more sampling rates is straightforward. Furthermore, we make the following contributions to fulfill the proposed scheme as presented in this paper:
\begin{itemize}
\item We design a new CNN structure for block up-sampling in the proposed scheme. To achieve higher reconstrution quality and simpler network structure, we explore a five-layer CNN for up-sampling, which features deconvolution of feature maps, multi-scale fusion, and residue learning. Moreover, we propose to use different networks for the up-sampling of luma and chroma components, respectively. The chroma up-sampling CNN reuses the luma information to improve its performance.
\item We investigate how to integrate the up-sampling CNN into the intra frame coding scheme. Besides allowing the encoder to choose the sampling rate for each CTU, as mentioned above, we also propose to allow the encoder to select the up-sampling method for each down-sampled CTU with selection from either CNN or fixed interpolation filters. To handle the boundary condition in block-wise up-sampling, we propose a two-stage up-sampling process where the first stage is within the block-by-block coding loop, and the second stage is out of the loop to refine the CTU boundaries. We also perform empirical study on how to decide the coding parameters of the down-sampled blocks to pursue frame-level rate-distortion optimization.
\item We perform extensive experiments to validate the proposed coding scheme as well as each proposed technique. The proposed scheme is implemented based on the HEVC reference software, and is shown to achieve significant bits saving compared with HEVC anchor especially at low bit rates. The proposed up-sampling CNN not only performs better, but also is simpler and computationally more efficient than the state-of-the-art image SR networks.
\end{itemize}

The remainder of this paper is organized as follows. In Section \ref{sec_previous_work}, we discuss related work on down/up-sampling-based coding and CNN-based image SR. Section \ref{sec_framework} presents the framework of the proposed block down/up-sampling-based coding scheme. The CNN structures for luma and chroma up-sampling are discussed in Section \ref{sec_CNN_Up_Samplng}. Coding parameters setting and the two-stage up-sampling process are elaborated in Sections \ref{sec_Coding_Params_Decision} and \ref{sec_CTU_Boundary_Refining}, respectively. Section \ref{sec_Experimental_Results} presents the experimental results, followed by conclusions in Section \ref{sec_Conclusion}.
Table \ref{Abbreviations_Explanation} lists the abbreviations used in this paper.
\begin{table}
\caption{List of Abbreviations}
\label{Abbreviations_Explanation}
\center
\begin{tabular}{l|l}
\hline
CNN                   & Convolutional Neural Network                      \\
\hline
CTU                   & Coding Tree Unit                                  \\
\hline
DCTIF                 & Discrete Cosine Transform based Interpolation Filter       \\
\hline
HEVC                  & High Efficiency Video Coding                      \\
\hline
HR                    & High-Resolution                                   \\
\hline
LR                    & Low-Resolution                                    \\
\hline
MSE                   & Mean-Squared-Error                                 \\
\hline
PSNR                  & Peak Signal-to-Noise Ratio                        \\
\hline
QP                    & Quantization Parameter                            \\
\hline
R-D                   & Rate-Distortion                                   \\
\hline
ReLU                  & Rectified Linear Unit \cite{nair2010rectified}                            \\
\hline
SR                    & Super-Resolution                                  \\
\hline
SRCNN                 & Super-Resolution Convolutional Neural Network \cite{dong2014learning}    \\
\hline
SSIM                  & Structural Similarity \cite{wang2004image}                            \\
\hline
UCID                  & Uncompressed Colour Image Database \cite{schaefer2004ucid}               \\
\hline
UHD                   & Ultra High Definition                             \\
\hline
VDSR                  & Very Deep network for Super Resolution \cite{kim2016accurate}           \\
\hline
\end{tabular}
\end{table}

\section{Related Work}
\label{sec_previous_work}
In this section, we review the previous work that relates to our research in two categories. The first is down/up-sampling-based image and video coding, and the second is recently emerging CNN-based image SR.
\subsection{Down/Up-sampling-Based Coding}
Down-sampling before encoding and up-sampling after decoding is a well known strategy for image and video coding in scenarios where the transmission bandwidth is limited. Many researches on this topic have been focused on developing efficient up-sampling methods. For example, the down/up-sampling-based video coding scheme in \cite{molina2006toward} adopts the video SR method proposed in \cite{segall2004bayesian}, which is specifically designed for compressed videos by incorporating information like motion vectors into the SR task using a Bayesian framework. The scheme proposed by Shen \emph{et al.} \cite{shen2011down} adopts another up-sampling method, which belongs to learning-based SR methods, and imposes constraints of nearest neighbor searching region and rectifies the ``unreal'' pixels using inter-resolution and inter-frame correlations. Another scheme proposed by Barreto \emph{et al.} \cite{barreto2007region} takes into account the locally variant image characteristics, and performs region-based SR to improve the reconstruction quality. The segmentation of regions is performed at the encoder side, and the segmentation map is signaled as side information to the decoder to guide the SR process.

The abovementioned researches all perform down-sampling of the entire image/frame. However, it is noted that a uniform down-sampling rate cannot suit for all the different image regions that have variant features. Locally adaptive down-sampling rates are then proposed. In \cite{nguyen2008adaptive}, the appropriate down-sampling rates have been derived through theoretical analyses. In \cite{lin2006adaptive}, compliant with block-based coding, down-sampling rates are made adaptive for each block and selected from $1 \times 1$, $1/2\times 1$, $1 \times 1/2$, and $1/2 \times 1/2$.

Most of the previous studies on down/up-sampling-based coding adopt fixed, hand-crafted interpolation filters for both down- and up-sampling. In this work, we propose to utilize CNN models for up-sampling to enhance the reconstruction quality. In addition, we also adopt block-level adaptive down-sampling rates with selection from $1 \times 1$ or $1/2 \times 1/2$, as extension to more down-sampling rates is straightforward.
\subsection{CNN for Image SR}
Super-resolution or resolution enhancement aims at reconstructing high-resolution (HR) signal from low-resolution (LR) observation, which has been studied extensively in the literature. Existing image SR methods can be categorized into interpolation-based, reconstruction-based, and learning-based ones \cite{yang2010image}. Recently, inspired by the success of deep learning, researchers have put more attention to learning-based SR using CNN.

Dong \emph{et al.} first proposed a CNN-based method for single image SR, termed SRCNN \cite{dong2014learning}, which has a simple network structure but demonstrated excellent performance.
Later on, several researches have been conducted to improve upon SRCNN at several aspects.
First, deeper networks have been explored to enhance the performance, such as the very deep network known as VDSR \cite{kim2016accurate}.
Second, it is observed that the training of SRCNN converges too slowly, and residue learning \cite{he2016deep}, i.e. learning the difference between LR and HR images rather than directly learning the HR images, is adopted to accelerate the training and also improves the reconstruction quality \cite{kim2016accurate}.
Third, the input to SRCNN is an interpolated version of LR image, which is to be enhanced by the network. The fixed interpolation filters before the network may not be optimal.
Thus, an end-to-end learning strategy, i.e. directly learning from the LR to the HR with embedding the resolution change into the network, is observed to perform better \cite{wang2016end}.

In this paper, we explore a new five-layer CNN structure for block up-sampling. Some key ingredients in the previously studied networks, such as residue learning and resolution change embedded in network, have been borrowed into our designed network. Our network structure is greatly simplified to reduce computational complexity, but still achieves satisfactory reconstruction quality, compared to the state-of-the-arts \cite{kim2016accurate,wang2016end}.

\section{Framework of the Proposed Scheme}
\label{sec_framework}
It is generally agreed that natural images/videos are equipped with locally variant features, and thus different regions may require different coding methods or parameters. For example, there are 35 intra prediction modes defined in HEVC intra coding, one of which can be selected for each block \cite{sullivan2012overview}. A down/up-sampling-based coding scheme provides more dimensions of freedom to explore so as to suit for different regions. While previous work has studied locally adaptive down-sampling rates \cite{lin2006adaptive,nguyen2008adaptive}, other dimensions such as adaptive down-sampling filters, adaptive coding parameters (e.g. quantization parameters), adaptive up-sampling filters, can be taken into account as well. Therefore, we propose to perform block-level down/up-sampling to embrace the flexibility, and to enable both adaptive down-sampling rates and adaptive up-sampling filters in the coding scheme. More adaptation will be considered in the future.
\begin{figure*}
\centering
\includegraphics[width=0.8\linewidth]{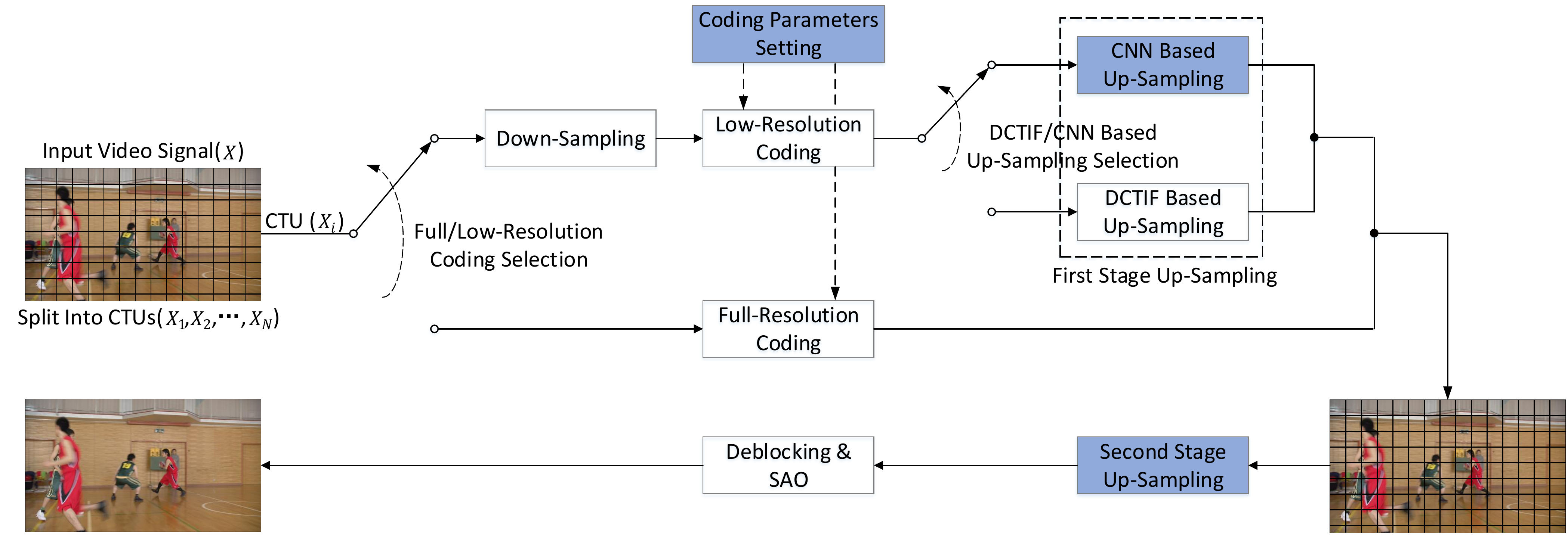}
\caption{The framework of our proposed intra frame coding scheme. The blue highlighted blocks indicate important modules in our scheme, which are discussed in detail in Sections \ref{sec_CNN_Up_Samplng}, \ref{sec_Coding_Params_Decision}, and \ref{sec_CTU_Boundary_Refining}, respectively. Note that both \emph{Full-Resolution Coding} and \emph{Low-Resolution Coding} are indeed intra coding (e.g. H.264 intra coding or HEVC intra coding), but working at different resolutions.}
\label{fig_framework}
\end{figure*}

Fig. \ref{fig_framework} depicts the flowchart of our proposed intra frame coding scheme. An input frame is divided into blocks while for each block the best coding mode is decided. In this paper, the block is chosen to be of the same size as CTU, i.e. consisting of 64$\times$64 luma samples (Y) and 2 channels of 32$\times$32 chroma samples (U and V, or Cb and Cr), due to the YUV 4:2:0 format. Each CTU can be either coded at its full resolution, or down-sampled and coded at low resolution. Here, the down-sampling is performed using the fixed filters presented in \cite{JCTVC-F158}.

Next, if the CTU is down-sampled and coded, it should be up-sampled back to its original resolution so as not to disrupt the normal intra coding of the subsequent CTUs. For this up-sampling, each down-sampled CTU can choose either CNN-based up-sampling, or the fixed, discrete cosine transform based interpolation filters (DCTIF) \cite{agbinya1992interpolation}. We adopt DCTIF in addition to our proposed CNN-based up-sampling, because DCTIF is already adopted in HEVC for fractional pixel interpolation for motion compensation \cite{sullivan2012overview}, and it is computationally simple but achieves good quality for smooth image regions. While CNN is much more complicated than DCTIF, we expect CNN to deal with complex image regions such as structures. The CNN-based up-sampling is elaborated in Section \ref{sec_CNN_Up_Samplng}.

There are two mode decision steps shown in Fig. \ref{fig_framework}. The first is for each down-sampled coded CTU, one up-sampling method is decided. This is performed by comparing the up-sampled results of both methods with the original CTU, and choosing the result with less distortion, since the down-sampled coding rate is the same. The second mode decision is to choose low-resolution coding or full-resolution coding for each CTU, which is performed by comparing the rate-distortion (R-D) costs of both coding modes. The distortion values of both coding modes are calculated at full resolution for fair comparison. Due to the down-sampling, low-resolution coding may incur much higher distortion but needs much less coding rate, thus it would be beneficial to adjust the coding parameters for down-sampled coded CTUs to pursue the overall R-D optimization, as elaborated in Section \ref{sec_Coding_Params_Decision}.

In addition, the block-level down/up-sampling bears a side effect of the boundary conditions during down- and up-sampling. Specifically, all the down- and up-sampling methods, including CNN-based ones, need appropriate boundary conditions. In general, such methods perform worse at image boundaries due to lack of information. We carefully address this problem. For down-sampling there are two cases: first, the original frame is entirely down-sampled to provide the down-sampled version of each CTU to be compressed; second, if a CTU chose full-resolution coding mode, the reconstructed CTU needs to be down-sampled so as to provide appropriate reference for the intra prediction of subsequent down-sampled CTUs. In both cases, we adopt the border replication method, i.e. replicating the values at the borders outwards, to provide the unavailable pixels at image boundaries or CTU boundaries. For up-sampling, we propose a two-stage method that uses different boundary conditions. The two-stage up-sampling is depicted in Fig. \ref{fig_framework}, and will be elaborated in Section \ref{sec_CTU_Boundary_Refining}.

\section{CNN-Based Up-sampling}
\label{sec_CNN_Up_Samplng}
Image SR is a severely ill-posed problem, and the key to relax the ill-posedness is the modeling of natural image prior. Training CNN for image SR is essentially embedding the natural image prior into the network parameters. And previous work \cite{dong2014learning,dong2016accelerating,wang2016end,kim2016accurate,kim2016deeply} has demonstrated that CNN-based SR outperforms almost all the other methods in terms of both objective and subjective reconstruction quality.
Hence, we hope to develop an efficient CNN-based up-sampling method to be applied into our intra frame coding scheme.

\begin{figure*}
\centering
\includegraphics[width=0.8\linewidth]{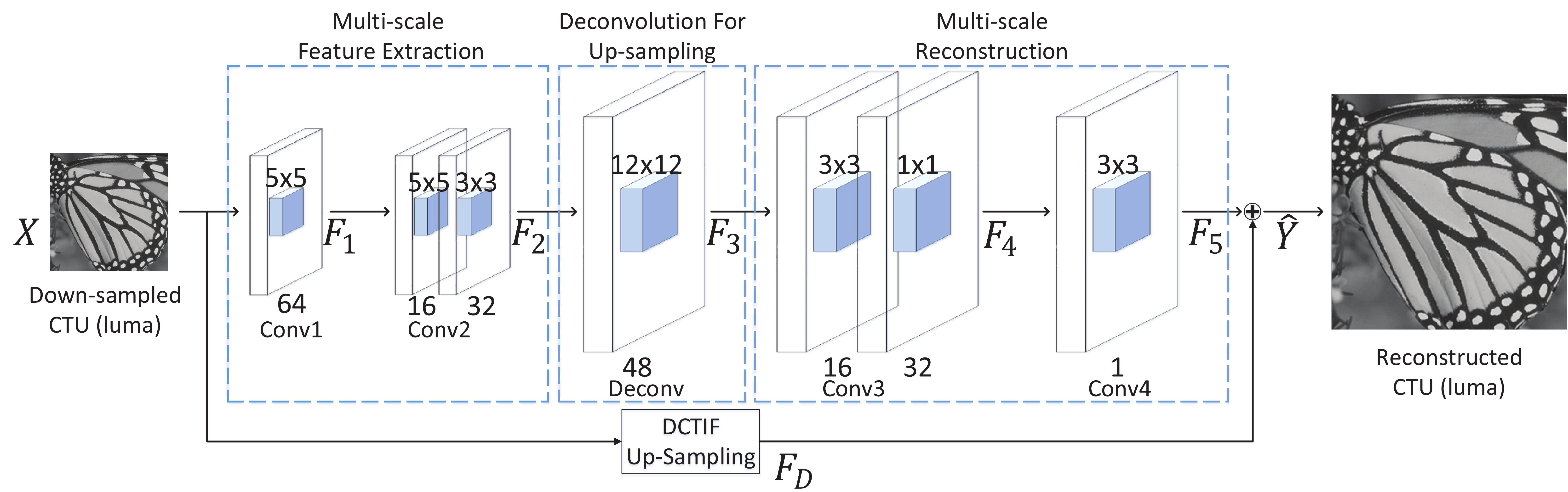}
\caption{Our designed five-layer CNN for the up-sampling of luma component. For each conv/deconv layer (e.g. Conv1), the numbers marked on the top (e.g. 5$\times$5) and on the bottom (e.g. 64) indicate its kernel size and the amount of channels of its output, respectively.}
\label{fig_luma_network}
\end{figure*}

\begin{figure*}
\centering
\includegraphics[width=0.8\linewidth]{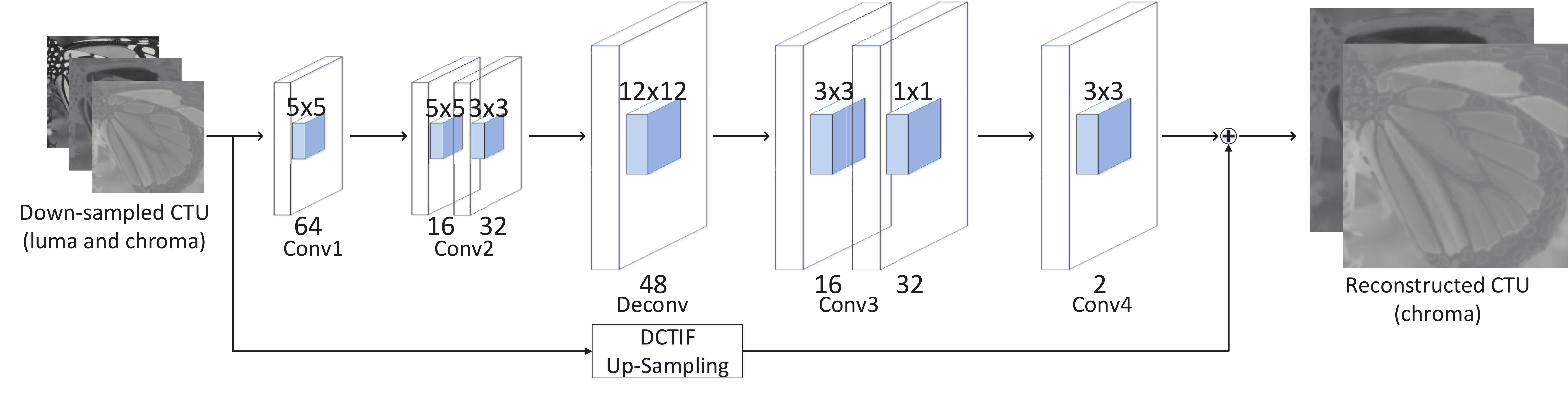}
\caption{Our designed CNN for the up-sampling of chroma components.}
\label{fig_chroma_network}
\end{figure*}

A trend in deep learning is to use deeper and deeper networks. For example, SRCNN \cite{dong2014learning} has 3 layers, but VDSR \cite{kim2016accurate} has 20 layers. Though the latter indeed achieves higher reconstruction quality, it also incurs higher computational cost. How to balance the reconstruction quality and computational complexity is an important issue to consider when designing the CNN structure, especially in video coding. In addition, note that the blocks to be up-sampled in our scheme have been compressed, and the distortion may be significant because of low bit rate coding. Thus, the CNN is expected to alleviate the distortion while at the same time to perform super-resolution. We are then motivated to explore a five-layer CNN for up-sampling, more complex than SRCNN (to deal with coding distortion) but much simpler than VDSR (to reduce computational cost). The network structures for the up-sampling of luma and chroma components are depicted in Figs. \ref{fig_luma_network} and \ref{fig_chroma_network}, and discussed in the following two subsections, respectively.
\subsection{CNN for Luma Up-sampling}
\label{sec_Luma_Network}
To achieve high reconstruction quality with a shallow network, we have borrowed some key ingredients from previous work, such as resolution change within the network, multi-scale fusion, and residue learning.
The CNN for luma up-sampling (shown in Fig. \ref{fig_luma_network}) can be divided into four parts: multi-scale feature extraction, deconvolution, multi-scale reconstruction, and residue learning, which are discussed one by one in the following.
\subsubsection{Multi-scale Feature Extraction}
\label{sec_Multi-scale_Feature_Extraction}
There are two layers designed to extract multi-scale features from the input LR block. Each layer consists of multiple convolutional kernels, each of which is followed by a rectified linear unit (ReLU) as nonlinear activation function.
It is well known that, an impressive advantage of CNN is to automate the feature extraction from raw data, which eliminates the necessity of hand-crafted features. Therefore, we directly input the LR compressed block into CNN without any pre-processing.

The first layer of CNN can be expressed as
\begin{equation}
\label{eqn_feature_extraction_layers}
\mathbf{F}_1(\mathbf{X}) = \max(0, \mathbf{W}_1*\mathbf{X}+\mathbf{B}_1)
\end{equation}
where $\mathbf{W}_1$ and $\mathbf{B}_1$ represent the convolutional filters and biases of the first layer, respectively, $\mathbf{X}$ is the input LR block, $\mathbf{F}_1$ indicates the feature maps of the first layer, and $*$ stands for convolution.

Since the input block is already compressed, it contains compression noise especially when quantization parameter (QP) is large. The feature maps extracted by the first layer may still contain noise, and thus the second layer is inserted to suppress noise and to enhance useful features:
\begin{equation}
\mathbf{F}_2(\mathbf{X})=
\begin{cases}
\mathbf{F}_{21}(\mathbf{X}) = \max(0, \mathbf{W}_{21}*\mathbf{F}_1(\mathbf{X})+\mathbf{B}_{21})\\
\mathbf{F}_{22}(\mathbf{X}) = \max(0, \mathbf{W}_{22}*\mathbf{F}_1(\mathbf{X})+\mathbf{B}_{22})
\end{cases}
\label{eqn_feature_enhancement_layers}
\end{equation}
where $(\mathbf{F}_{21},\mathbf{F}_{22})$, $(\mathbf{W}_{21}, \mathbf{W}_{22})$, $(\mathbf{B}_{21}, \mathbf{B}_{22})$ are the extracted feature maps, convolutional filters, and biases, respectively. Note that there are two sets of convolutional kernels that have different kernel sizes in the second layer. Different sized kernels have receptive fields at different scales, and the combination of them is capable in effectively aggregating multi-scale information, which has been widely adopted in computer vision \cite{serre2007robust,szegedy2015going}. Here in the second layer, the combination of different sized kernels provides multi-scale features to be explored for super-resolution. Note that the output feature maps $\mathbf{F}_{21}$ and $\mathbf{F}_{22}$ are directly concatenated and fed into the next layer.

\subsubsection{Deconvolution}
\label{sec_Deconvolution_For_Up-Sampling}
In most of the previous work on image SR, either CNN-based or not, an input LR image is first up-sampled by a fixed interpolation filter (e.g. bicubic) and then enhanced. The enhancement process does not change the resolution. However, it has been pointed out that the fixed interpolation filter before enhancement may cause the loss of important information in the original LR image. An end-to-end learning, embedding the resolution change into CNN, is believed better \cite{wang2016end}. There are two techniques in CNN for resolution upgrade: un-pooling \cite{dosovitskiy2015learning} and deconvolution \cite{noh2015learning}. While the un-pooling tends to yield enlarged but sparse output, we adopt the deconvolution in our designed CNN.

As shown in Fig. \ref{fig_luma_network}, the third layer performs deconvolution of the multi-scale feature maps extracted by the second layer. Deconvolution changes the resolution of input by multiplying each input pixel by a filter to produce a window, and then summing over the resulting windows. A ReLU is then appended to the deconvolution, leading to
\begin{equation}
\mathbf{F}_3(\mathbf{X})= \max(0, \mathbf{W}_3\star \mathbf{F}_2(\mathbf{X})+\mathbf{B}_3)
\label{eqn_deconvolution__layers}
\end{equation}
where the symbol $\star$ denotes deconvolution.

The relative position of the deconvolution layer in the CNN is also an issue to consider. It can be put at the beginning, in the middle, or at the end of the entire CNN. In our designed CNN, the deconvolution layer is used to enlarge the multi-scale feature maps and the enlarged features are then used to reconstruct HR image, then it is in the middle. We have tried to put it at other positions, but empirical results show the decrease of reconstruction quality then.

\subsubsection{Multi-scale Reconstruction}
\label{sec_Multi_scale_Reconstruction}
The reconstruction stage is composed by two convolutional layers.
The fourth layer, similar to the second, performs multi-scale fusion by using two sets of convolutional kernels with different sizes,
\begin{equation}
\mathbf{F}_4(\mathbf{X})=
\begin{cases}
\mathbf{F}_{41}(\mathbf{X}) = \max(0, \mathbf{W}_{41}*\mathbf{F}_3(\mathbf{X})+\mathbf{B}_{41})\\
\mathbf{F}_{42}(\mathbf{X}) = \max(0, \mathbf{W}_{42}*\mathbf{F}_3(\mathbf{X})+\mathbf{B}_{42})
\end{cases}
\label{eqn_1st_reconstruction_layer}
\end{equation}
This layer takes into account both long- and short-range contextual information for reconstruction.

Then, the fifth layer performs reconstruction,
\begin{equation}
\mathbf{F}_5(\mathbf{X}) = \mathbf{W}_5*\mathbf{F}_4(\mathbf{X})+\mathbf{B}_5
\label{eqn_2nd_reconstruction_layer}
\end{equation}
Note that the fifth layer has no nonlinear unit.
\subsubsection{Residue Learning}
\label{sec_Residue_Learning}
Residue learning in CNN is proposed by He \emph{et al.}, who introduced skip-layer connections in CNN to achieve both faster convergence in training and better performance \cite{he2016deep}. We also adopt residue learning in our network and have observed indeed faster convergence in training. Specifically, the down-sampled block is up-sampled by a fixed interpolation filter (DCTIF in this paper for consistency) and then added to the reconstruction produced by the five-layer CNN,
\begin{equation}
\mathbf{F}_D(\mathbf{X}) = \mathrm{DCTIF}(\mathbf{X})
\label{eqn_DCTIF_Up_Sampling}
\end{equation}
\begin{equation}
\hat{\mathbf{Y}}(\mathbf{X}) = \mathbf{F}_5(\mathbf{X}) + \mathbf{F}_D(\mathbf{X})
\label{eqn_Reconstruction}
\end{equation}
In other words, the five-layer CNN is supposed to learn the difference between an original block and its degraded version, where the degraded version is generated by down-sampling the block, coding, and then up-sampling by DCTIF. The difference is indeed the high-frequency details in the original block that have been lost during down-sampling and coding. Learning to recover high-frequency details instead of the original image is a common strategy in image SR, with or without CNN \cite{kim2016accurate,sun2003image}.

Let the original HR block be $\mathbf{Y}$, the difference between $\mathbf{Y}$ and $\hat{\mathbf{Y}}(\mathbf{X})$ calculated by mean-squared-error (MSE) drives the training of our CNN. The MSE is minimized by means of stochastic gradient descent together with standard error back-propagation algorithm.

\subsection{CNN for Chroma Up-sampling}
\label{sec_Chroma_Network}
In most of the previous work on image SR, chroma components are simply interpolated by a fixed filter (e.g. bicubic) without enhancement. This is because human vision tends to be less sensitive to the change of chrominance signal, which is also the reason why the chroma components have a lower resolution in YUV 4:2:0 format. However in our coding scheme, we may further down-sample the chroma components and need to up-sample them, so we have designed a separate CNN for chroma up-sampling to achieve higher reconstruction quality. The chroma up-sampling CNN is depicted in Fig. \ref{fig_chroma_network}, whose structure is quite similar to the luma one but augmented with two features:
        \begin{figure}
        \centering
        \includegraphics[width=\linewidth]{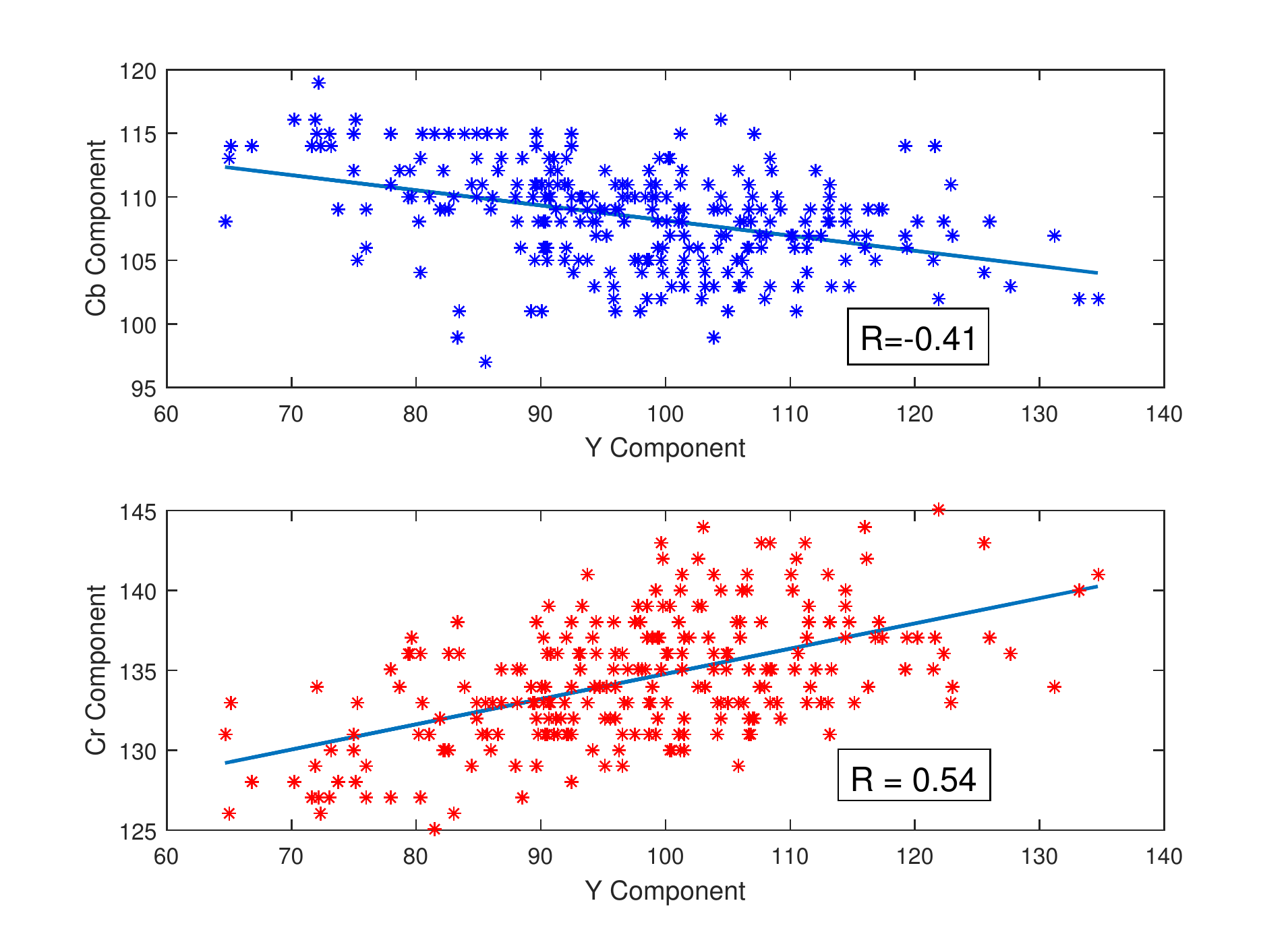}
        \caption{Example scatter plots showing the correlations between different channels of video. The data used in these plots come from a $32\times 32$ block of the \emph{Cactus} sequence. Correlation coefficient (R) is shown inside the plots.}
        \label{fig_Correlation_YUV}
        \end{figure}

\subsubsection{Incorporating Luma Information}
In the widely adopted YUV 4:2:0 format, luma and chroma components have been decomposed by conversion from RGB to YCbCr in advance. However, the decomposition did not fully remove the correlation among the three channels of RGB. There is still correlation between Y and Cb/Cr as can be observed from the example plots in Fig. \ref{fig_Correlation_YUV}. Motivated by this, predicting chroma from luma has been proposed for video coding \cite{lee2009intra,zhang2014chroma}. Similarly in this paper, we incorporate the luma information during the up-sampling of chroma components to improve the reconstruction quality. Moreover, the correlation between Y and Cb/Cr cannot be well described by simple linear models, as shown in Fig. \ref{fig_Correlation_YUV}, which inspires us to leverage the non-linear CNN models to exploit such correlation.

As shown in Fig. \ref{fig_chroma_network}, we use all the three channels (Y, Cb, and Cr) as input to CNN. Note that for down-sampled CTUs, the luma component has 32$\times$32 elements while the chroma components have only 16$\times$16 elements each. We further down-sample the luma component to the same size as chroma to simplify the network design. Then, cross-channel features can be extracted by the first layer, and processed by the following layers sequentially.

\subsubsection{Joint Training of Cb and Cr}
While it is possible to train two separate networks for Cb and Cr respectively, we believe the high similarity between Cb and Cr can help reduce the amount of required models. Specifically, the CNN shown in Fig. \ref{fig_chroma_network} outputs reconstructed Cb and Cr simultaneously, i.e. the former four layers are exactly the same for Cb and Cr, and only the last layer is different. During training, the MSE of both Cb and Cr is used as the objective of minimization. This design leads to fewer trained models, while incurs negligible loss of reconstruction quality, as observed from our empirical results.

\section{Coding Parameters Setting}
\label{sec_Coding_Params_Decision}
\begin{figure*}
\centering
\includegraphics[width=2.2in]{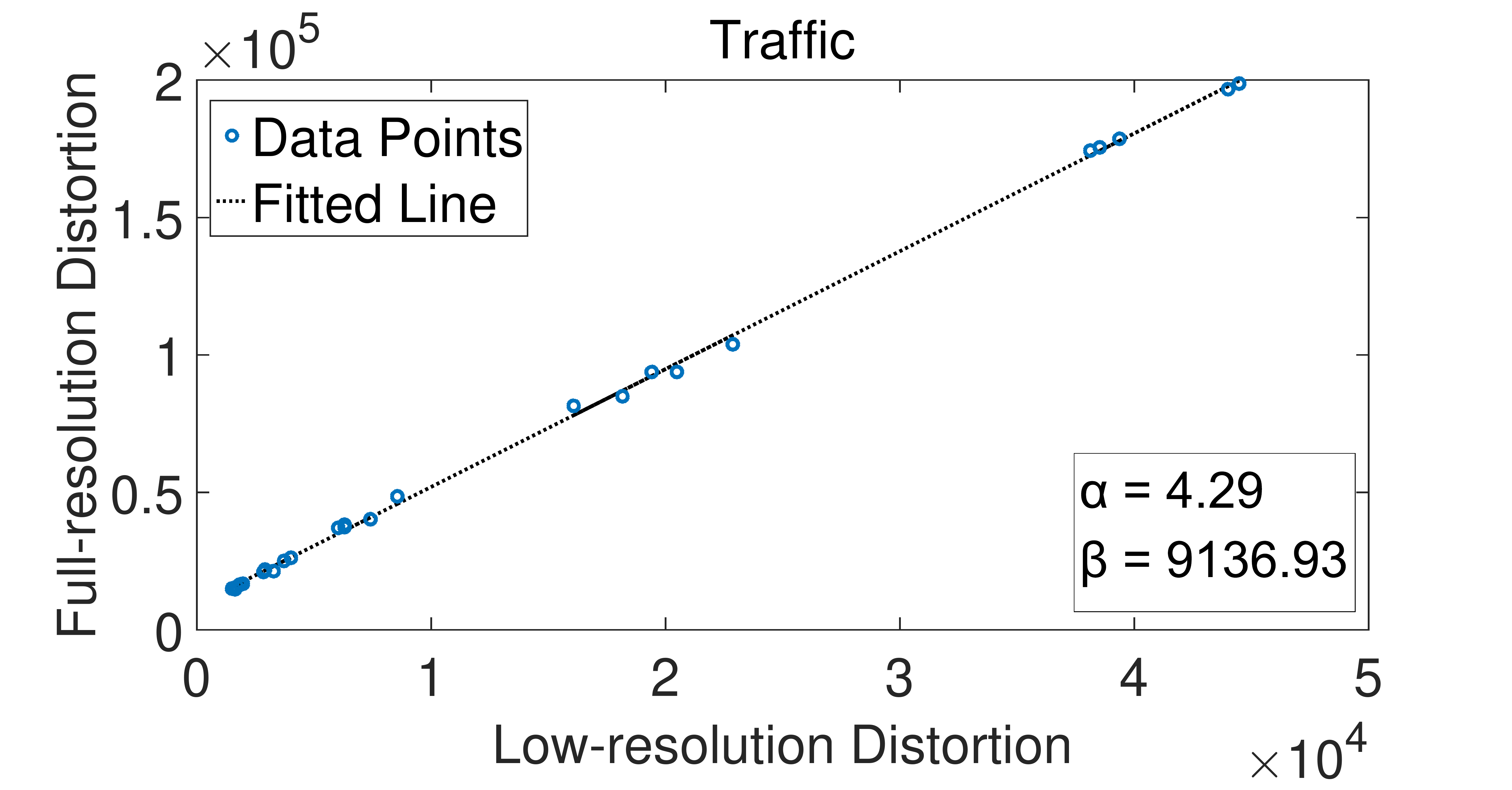}
\includegraphics[width=2.2in]{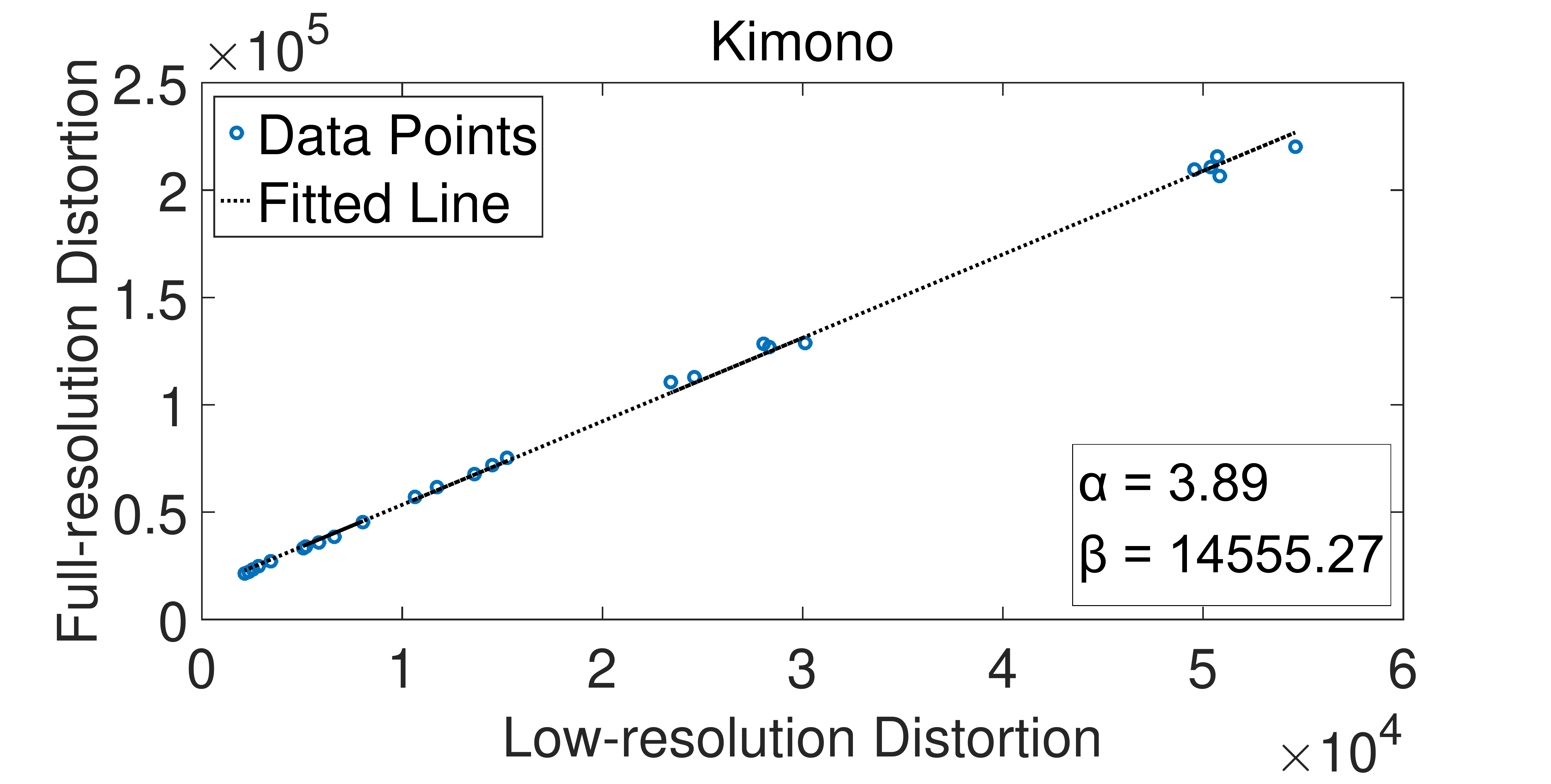}
\\
\includegraphics[width=2.2in]{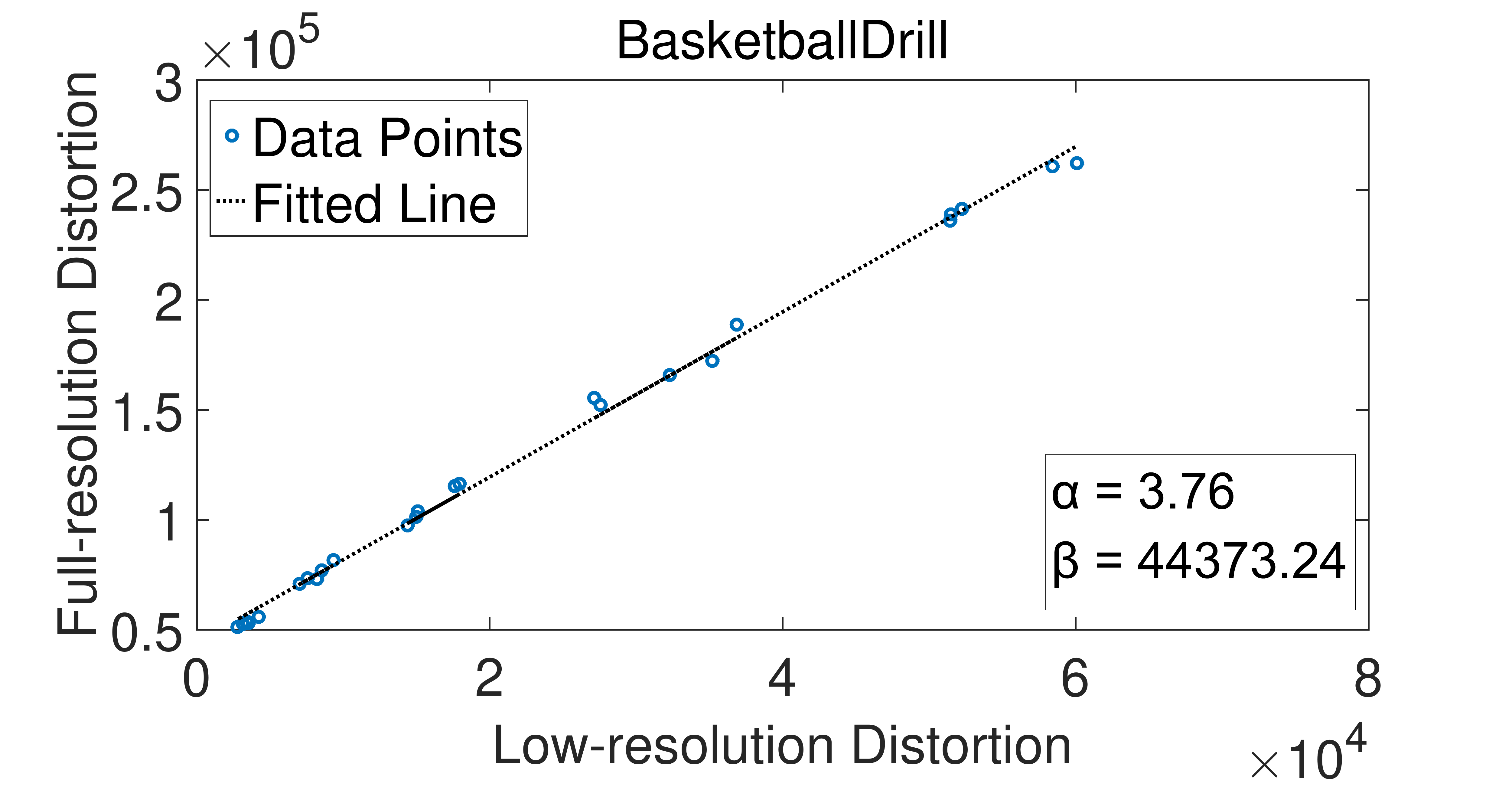}
\includegraphics[width=2.2in]{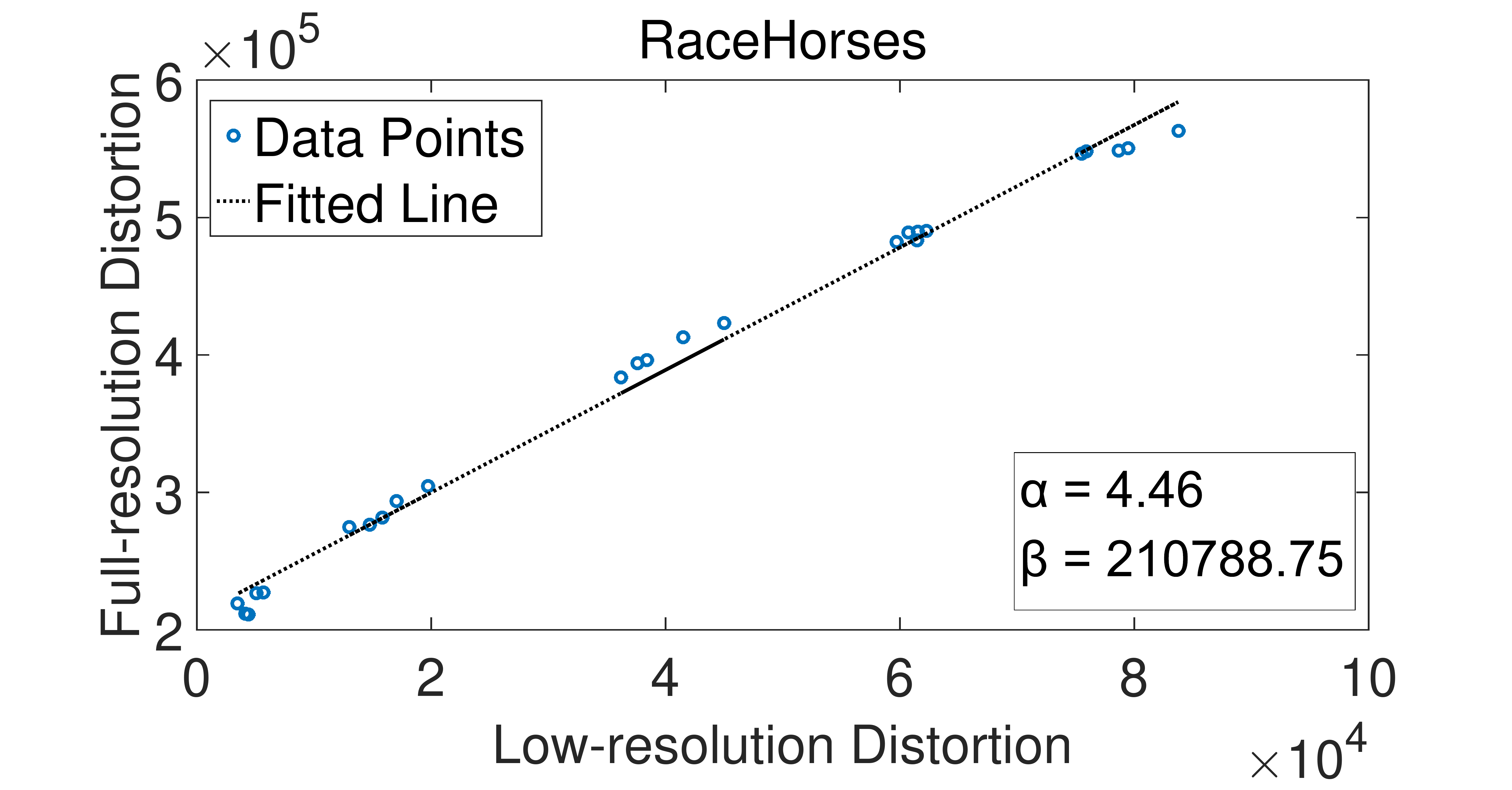}
\caption{Example plots showing the relation between the distortion calculated at full resolution and at low resolution. The data used in these plots come from 4 CTUs selected from 4 sequences indicated in the plots. Linear fitting coefficients ($\alpha$ and $\beta$) are shown inside the plots.}
\label{fig_D_D0}
\end{figure*}
In this section, we would like to derive the optimal coding parameters for down-sampled CTUs so as to pursue frame-level R-D optimization. We start from the basic objective function of R-D optimization, i.e.
\begin{equation}
\label{eqn_org_objective_function}
J = \sum_{i = 1}^N{D_{i}} + \lambda \sum_{i = 1}^N{R_{i}}
\end{equation}
where $J$ is the overall R-D cost, $D_{i}$ and $R_{i}$ are the distortion and rate of the $i$-th CTU, respectively, and $N$ is the total number of CTUs in the frame. $\lambda$ is the Lagrangian multiplier.
In the case of intra frame coding, the compression of each CTU can be regarded as approximately independent, because of the less accurate intra prediction between CTUs \cite{li2016lambda}.
Therefore, we consider the R-D cost of one CTU, and for simplicity the subscript $i$ is omitted hereafter.

In our coding scheme, the CTU can be coded at full resolution or at low resolution, but in both coding modes, the distortion $D$ shall be calculated at full resolution. However, during low-resolution coding, it is not easy to calculate the full-resolution distortion, denoted by $D_{full}$. Specifically, the down-sampled CTU (32$\times$32 in luma) is compressed by normal HEVC intra coding, during which the quadtree partition, the intra prediction modes, the quantized transform coefficients, as well as other syntax elements, need to be determined in an R-D optimized fashion. If $D_{full}$ is requested in low-resolution coding, then the down-sampled CTU needs to be up-sampled many times during the R-D optimization process of low-resolution coding. It is not only computationally expensive, but also not friendly to up-sampling that, as mentioned before, is sensitive to the lack of proper boundary conditions. Therefore, we prefer calculating the distortion directly at the low resolution, i.e. $D_{low}$, during low-resolution coding. Accordingly, in the low-resolution coding mode, $D_{full}$ is calculated only once, after the down-sampled CTU is entirely compressed and up-sampled.

Here, we take an empirical approach for investigation of the relation between $D_{full}$ and $D_{low}$. We have compressed many natural images/videos using the low resolution coding mode and different QPs, and calculated the pairs of $(D_{full},D_{low})$ in terms of sum-of-squared-difference. Some typical results are shown in Fig. \ref{fig_D_D0}, indicating that a linear model can be used to describe the relation, i.e.
\begin{equation}
D_{full} = \alpha \times D_{low} + \beta
\label{eqn_D_D0}
\end{equation}
The fitted values of $\alpha$ and $\beta$ are also shown in Fig. \ref{fig_D_D0}. Note that different CTUs have different values. This equation seems quite intuitive, as the full-resolution distortion can be decomposed into two parts, one part incurred by the low-resolution coding, and the other part corresponding to the lost high-frequency information during down-sampling.

Given (\ref{eqn_D_D0}), the R-D cost of one CTU can be written as
\begin{equation}
\begin{array}{rl}
\label{eqn_RD_Sub1}
J & = D_{full} + \lambda  R \\
            & = \alpha  D_{low} + \beta + \lambda  R \\
            & = \alpha( D_{low} + \frac{\lambda}{\alpha}  R ) + \beta
\end{array}
\end{equation}
The R-D cost during low-resolution coding can be written as
\begin{equation}
\label{eqn_RD_Sub2}
J_{low} = D_{low} + \lambda_{low}  R
\end{equation}
Note that the $R$ is the same in both (\ref{eqn_RD_Sub1}) and (\ref{eqn_RD_Sub2}). Thus, if we choose $\lambda_{low}=\frac{\lambda}{\alpha}$, then the optimization of (\ref{eqn_RD_Sub2}) and that of (\ref{eqn_RD_Sub1}) are equivalent. Moreover, in HEVC the quantization parameter (QP) is known to depend on the Lagrangian parameter $\lambda$, i.e.
\begin{equation}
\label{eqn_lambda_QP_Overall}
\lambda = c \times 2^{\frac{QP-12}{3}}
\end{equation}
Then, the QP during low-resolution coding should be changed accordingly into
\begin{equation}
\label{eqn_QP}
QP_{low} = QP - 3 \times \log_{2}{\alpha}
\end{equation}
This equation is also intuitively meaningful, because the low-resolution coding in general leads to less rate but more distortion, and we need to lower the QP to make both rate and distortion of low-resolution coding to be comparable to that of full-resolution coding.

\begin{figure}
\centering
\includegraphics[width=0.8\linewidth]{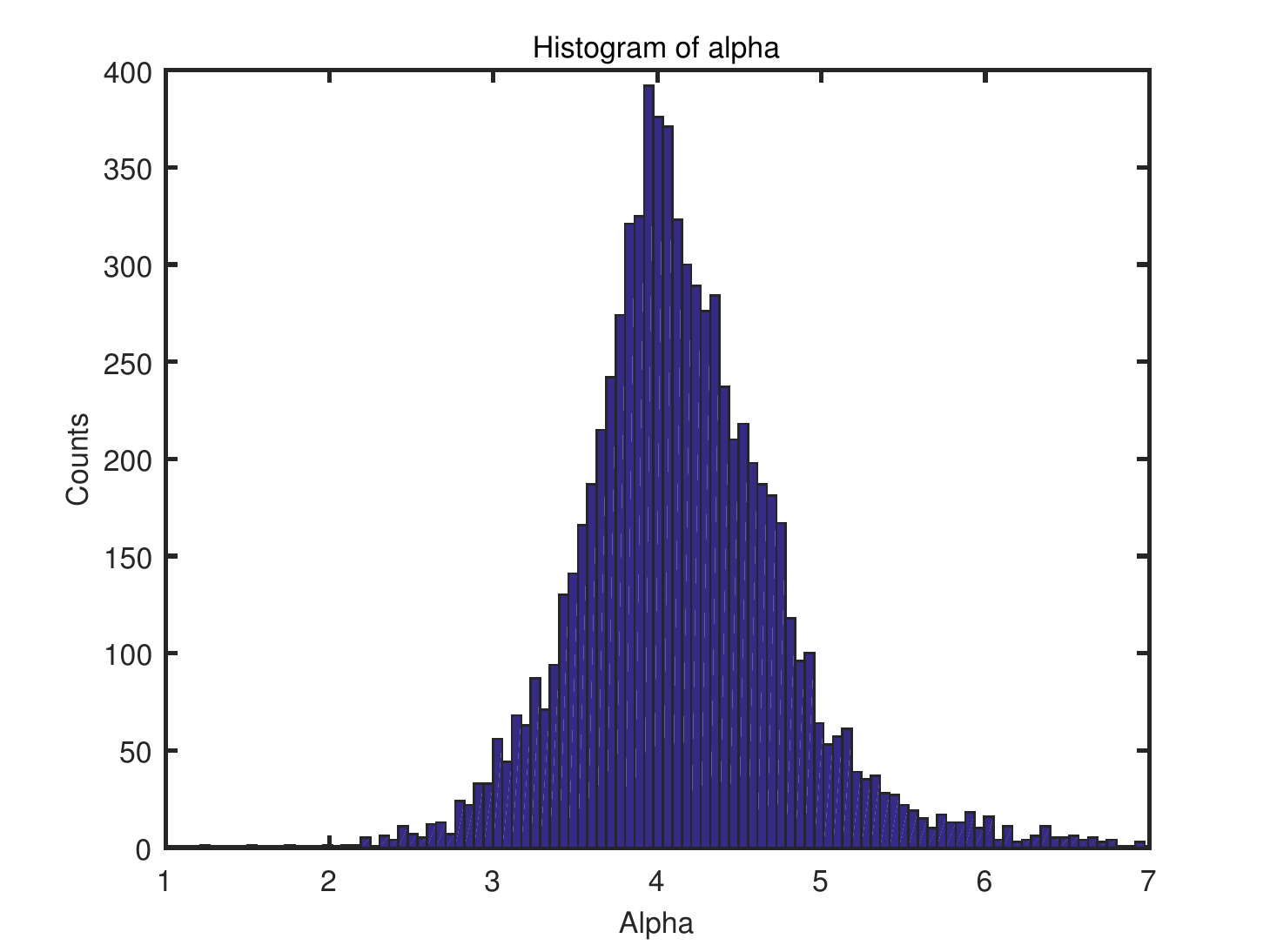}
\caption{The distribution of $\alpha$ for all the CTUs of all the test sequences.}
\label{fig_Histogram}
\end{figure}

However, if we adjust QP according to (\ref{eqn_QP}), the $\alpha$ value of each CTU is distinct, i.e. each low-resolution coded CTU has a different QP, which requires additional bits to encode. Besides, it is not easy to determine the $\alpha$ value of each CTU in practice. We are then motivated to use a predefined $\alpha$ or equivalently a fixed delta QP for low-resolution coding. To this end, we perform statistical analysis of the fitted $\alpha$ values using many natural images/videos. The empirical distribution of $\alpha$ is plotted in Fig. \ref{fig_Histogram}, indicating the mode of $\alpha$ is around 4. This number is reasonable as our down-sampling rate is $1/2\times 1/2$. Therefore, in our experiments, we set fixed coding parameters for low-resolution coding, i.e.
\begin{equation}
\label{eqn_RD_Specific}
\lambda_{low} = \lambda/4
\end{equation}
\begin{equation}
\label{eqn_QP_Specific}
QP_{low} = QP - 6
\end{equation}

\section{Two-Stage Up-sampling}
\label{sec_CTU_Boundary_Refining}
\begin{figure}
\centering
\includegraphics[width=0.8\linewidth]{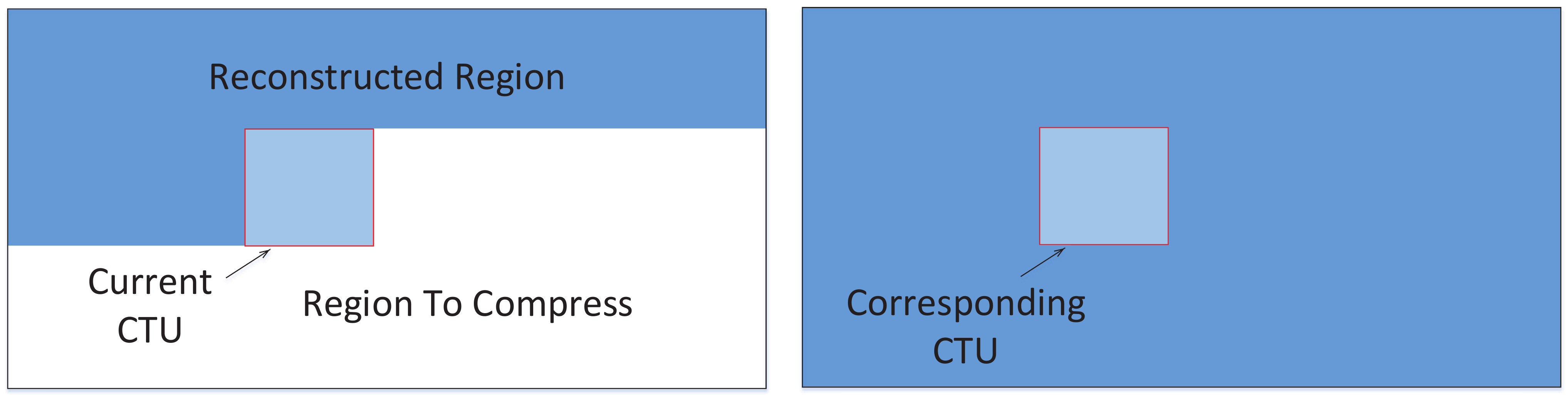}
\caption{The two stages of block up-sampling utilize different boundary conditions. Left: For the first stage, bottom and right boundaries are not available during up-sampling. Right: For the second stage, all boundaries are available for up-sampling.}
\label{fig_refining_block_boundary}
\end{figure}

We design a two-stage up-sampling process as shown in Fig.~\ref{fig_framework}. The difference between two stages can be observed from Fig. \ref{fig_refining_block_boundary}. In the first stage, the CTU needs to be up-sampled for the coding of subsequent CTUs, the up-sampling at this stage can use the top and left boundaries but cannot use the bottom and right ones as they are not compressed yet. In our implementation, we fill the unavailable boundaries with zero values. However, in the second stage, the entire frame has been compressed, so the up-sampling can use all available boundaries. In essence, the second stage refines the region of each up-sampled CTU around its bottom and right boundaries. This is valid for both CNN- and DCTIF-based up-sampling.

The second stage of up-sampling is performed for only the CTUs that have chosen the low-resolution coding mode, and the up-sampling method (CNN-based or DCTIF) is already decided in the first stage. The up-sampling result of the second stage just replaces that of the first stage. The same process is performed at both encoder and decoder, then no overhead bit is required.

\section{Experimental Results}
\label{sec_Experimental_Results}
We conduct extensive experiments to evaluate the performance of the proposed methods.
Experimental settings are introduced, followed by the detailed experimental results and analyses in this section.

\subsection{Experimental Settings}
\label{sec_Experimental_Settings}
\subsubsection{Implementation and Configuration}
\label{sec_software}
We have implemented our proposed intra frame coding scheme based on the reference software of HEVC, i.e. HM version 12.1\footnote{\url{https://hevc.hhi.fraunhofer.de/svn/svn_HEVCSoftware/tags/HM-12.1/}}. In HEVC intra coding, each CTU is partitioned into coding units based on a quadtree, and the luma and chroma components of one CTU must follow the same quadtree. To comply with this, the mode decision between full- and low-resolution coding is performed at CTU level combining luma and chroma, i.e. the R-D costs of luma and chroma are summed up to make decision. On the contrary, the mode decision of which up-sampling method is performed individually for Y, Cb, and Cr, i.e. if a CTU chooses low-resolution coding, three binary flags are required to indicate CNN-based or DCTIF for the channels Y, Cb, and Cr, respectively.

The CNN-based up-sampling method has been realized using Caffe \cite{jia2014caffe}, a popular framework for deep learning, to reuse its highly efficient implementation of convolutions.

We use the all-intra configuration suggested by HEVC common test conditions \cite{bossen2011common}.
Considering down/up-sampling-based coding is a useful tool especially at low bit rates, the QP is set to \{32, 37, 42, 47\}.
BD-rate \cite{bjontegaard2001calcuation} is adopted to evaluate the compression efficiency, where for the quality metric we use both PSNR and structural similarity (SSIM) \cite{wang2004image}, as the latter is believed to be more consistent with subjective quality.
\subsubsection{Test Sequences}
\label{sec_test_sequences}
The HEVC common test sequences, including 20 video sequences of different resolutions known as Classes A, B, C, D, E \cite{bossen2011common}, are used for experiments. Class F (screen content videos) is excluded as our proposed technique is designed for natural videos. In addition, to demonstrate the performance on high definition videos, we use five 4K (3840$\times$2160) sequences from the SJTU dataset \cite{song2013sjtu} in experiments, as shown in Table \ref{tab_all_seq}. For each sequence, we use only the first frame in experiments, and our empirical results indicate that the comparative results using entire sequences have similar trends.
\begin{table}
  \centering
  \caption{Characteristics of the UHD Test Sequences}
    \begin{tabular}{l|c|l|c}
    \hline
    Source & Resolution & Name & Frame Rate \\
    \hline
   \multirow{5}{*}{SJTU UHD}     & \multirow{5}{*}{3840$\times$2160} & Fountains      &  \multirow{5}{*}{30 fps} \\
          &      & Runners        &     \\
          &      & Rushhour       &     \\
          &      & TrafficFlow    &     \\
          &      & CampfireParty  &     \\
    \hline
    \end{tabular}%
  \label{tab_all_seq}%
\end{table}%
\subsubsection{CNN Training}
\label{sec_training}
The Caffe software is also used to train CNN models. We use the Uncompressed Colour Image Database (UCID) \cite{schaefer2004ucid}, which consists of 1338 natural images, to prepare the training data. The training data and test data (video sequences) have no overlap to demonstrate the generalization ability of CNN. The images in UCID are compressed by our scheme using different QPs, but all CTUs are forced to use the low-resolution coding mode and DCTIF for up-sampling. The reconstructed LR CTUs together with the original ones are formed into pairs of $(\mathbf{X},\mathbf{Y})$ to train the CNN as described in Section \ref{sec_CNN_Up_Samplng}. It is worth noting that we have trained a different model for each QP and for Y or Cb/Cr, so in total we have 8 CNN models corresponding to the four QPs.
\subsection{Results and Analyses}
\label{sec_Experimental_Analysis}
\subsubsection{Overall Performance}
\label{sec_RD_Performance Analysis}
The overall performance measured by BD-rate is shown in Table \ref{Results_of_All_Sequences}.
Columns under ``Anchored on HEVC'' are the results comparing our scheme with HM 12.1 anchor.
As can be observed, our scheme improves the coding efficiency significantly, leading to on average 5.5\%, 6.0\%, and 2.2\% BD-rate reductions on Y, U, and V, respectively, for HEVC test sequences (Classes A--E).
As for UHD test sequences, our scheme achieves even higher coding gain, i.e. 9.0\%, 1.6\%, and 3.2\% BD-rate reductions on Y, U, and V.

It is worth noting that the images used in training all have a bit-depth of 8, but there are two 10-bit sequences for test, i.e. Nebuta and SteamLocomotive (in Class A). For these two sequences, the BD-rate reduction on Y is limited but on U and V are still significant. It is possible to further improve for such sequences by including high-dynamic-range images during training.

For a few sequences, we observe the BD-rate on U and V is positive, indicating performance loss of our scheme, but for such sequences the BD-rate reduction on Y is still visible. The reason of such phenomenon is that for several CTUs, the luma component prefers low-resolution coding but the chroma components prefer full-resolution coding. However, our current implementation forces the modes (full or low) of luma and chroma to be the same to suit for HEVC intra coding. This constraint may be removed in the future to pursue better performance.

In addition, when using SSIM as quality metric, the BD-rate reductions are more significant, i.e. 8.8\% and 10.5\% on Y for HEVC and UHD sequences, respectively. Thus, we believe down/up-sampling-based coding is more friendly to the subjective quality at low bit rates.

We conduct another experiment to demonstrate the benefit of using CNN for up-sampling in addition to the fixed interpolation filters. In this experiment, the CNN-based up-sampling in our scheme is disabled and DCTIF is the only up-sampling method. Comparative results measured by BD-rate are presented in columns under ``Anchored on HEVC+DCTIF'' in Table \ref{Results_of_All_Sequences}.
As can be observed, adopting CNN-based up-sampling improves the coding efficiency of down/up-sampling-based coding by a considerable margin.
The BD-rate reductions on Y, U, and V are on average 4.3\%, 10.0\%, and 6.0\% for HEVC test sequences, and on average 5.1\%, 10.5\%, and 9.9\% for UHD test sequences.

\begin{figure*}
\centering
\subfigure[ ]
{
\includegraphics[width=0.3\linewidth]{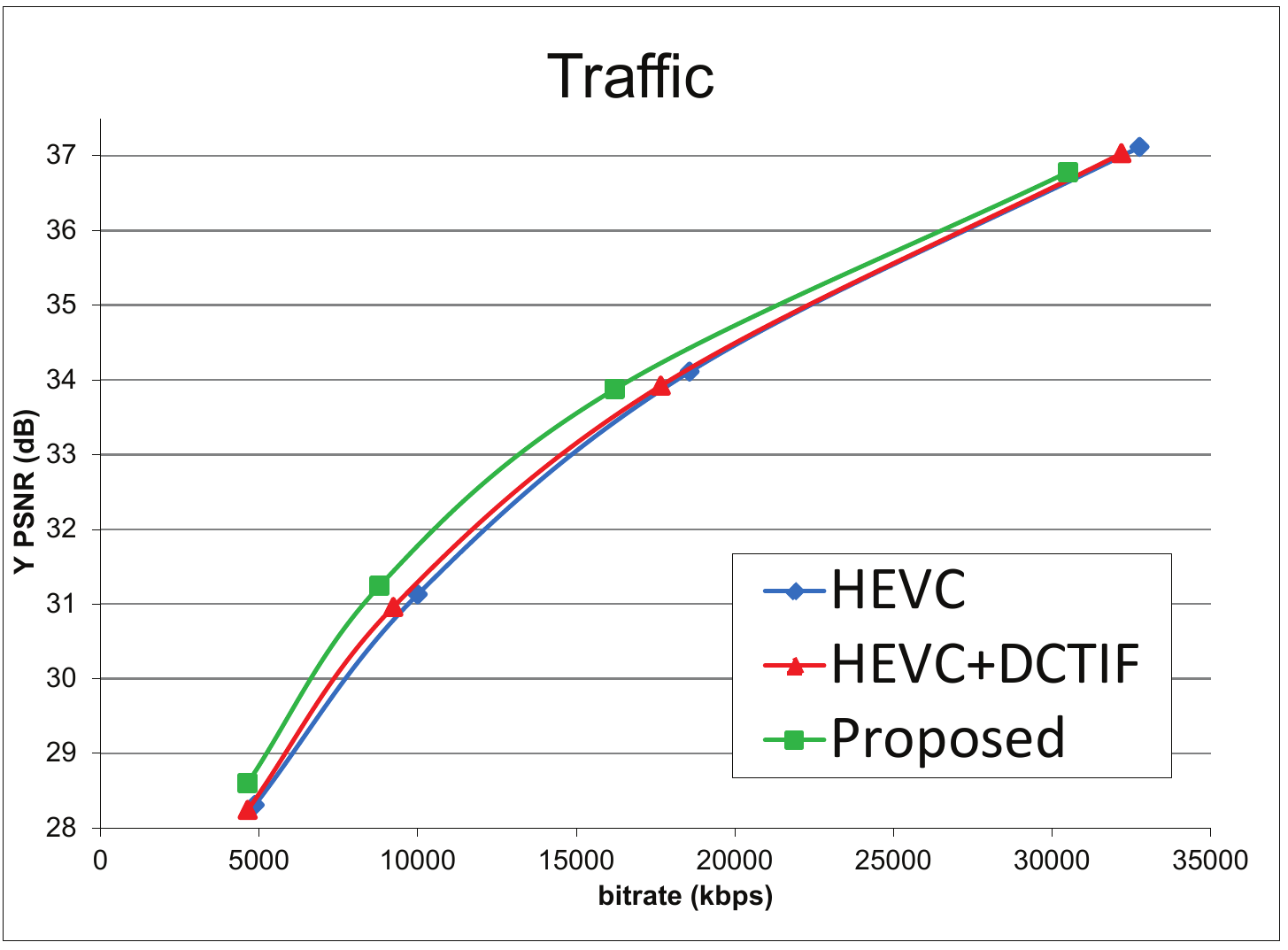}
}
\subfigure[ ]
{
\includegraphics[width=0.3\linewidth]{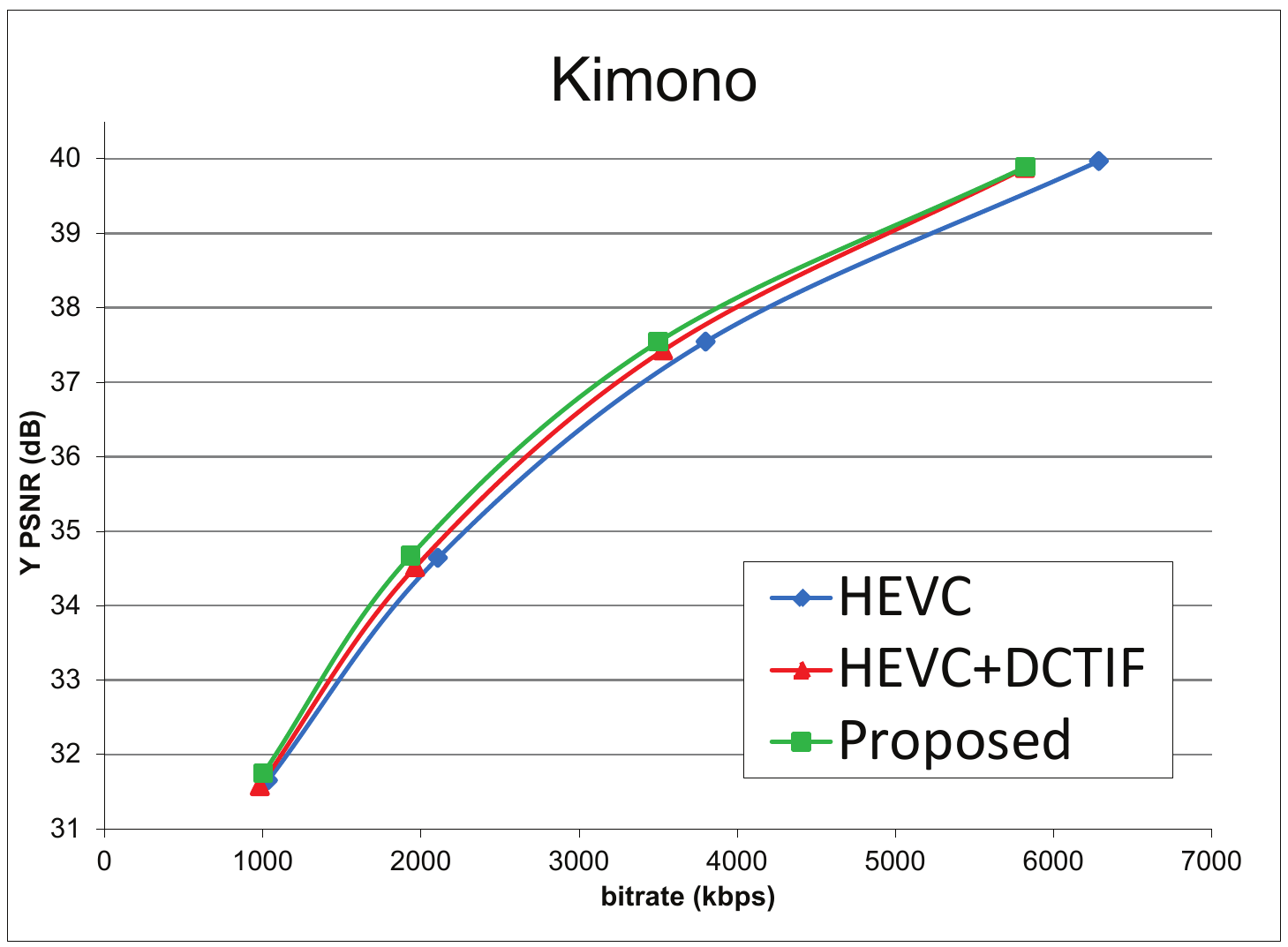}
}
\subfigure[ ]
{
\includegraphics[width=0.3\linewidth]{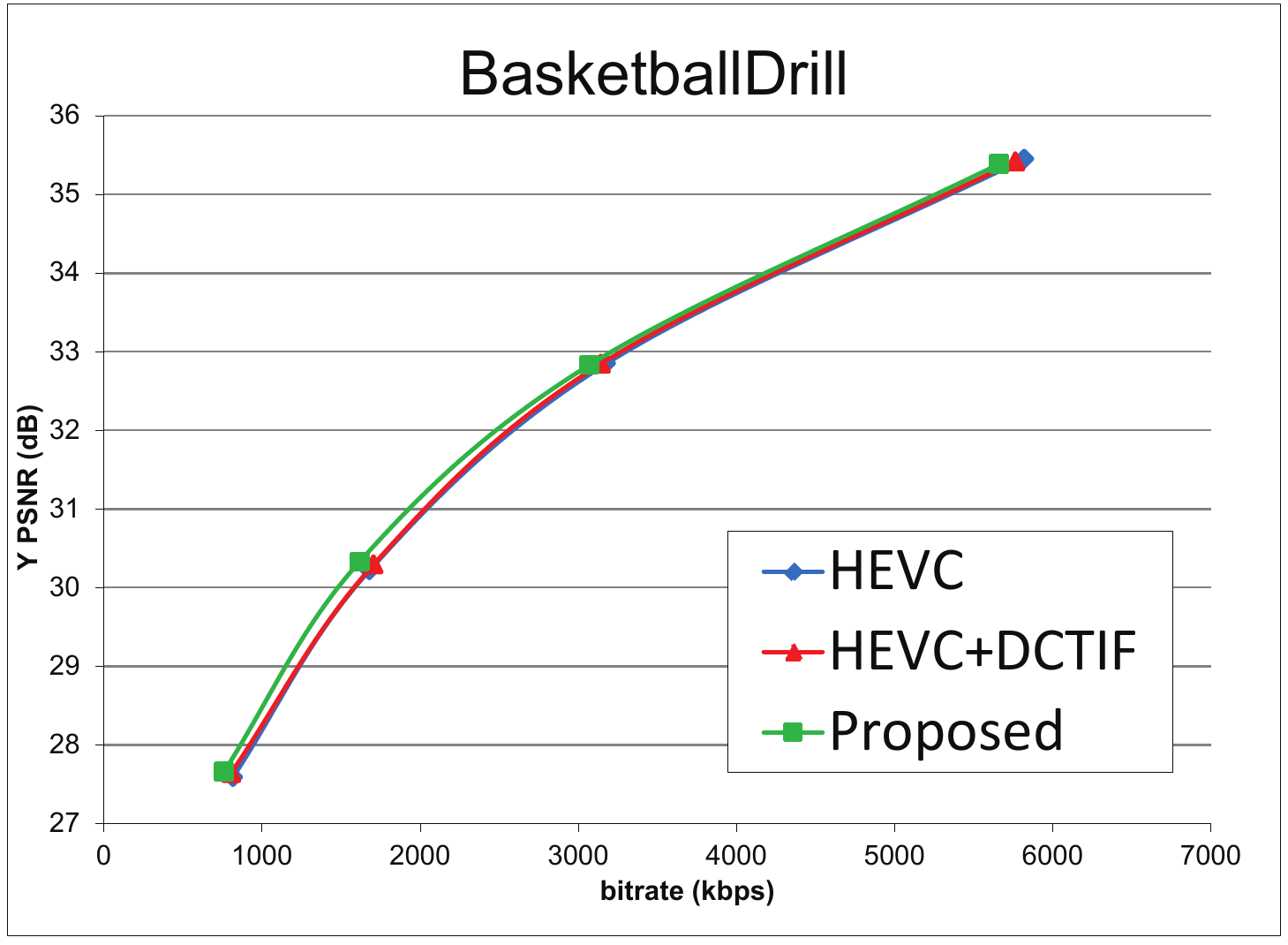}
}
\subfigure[ ]
{
\includegraphics[width=0.3\linewidth]{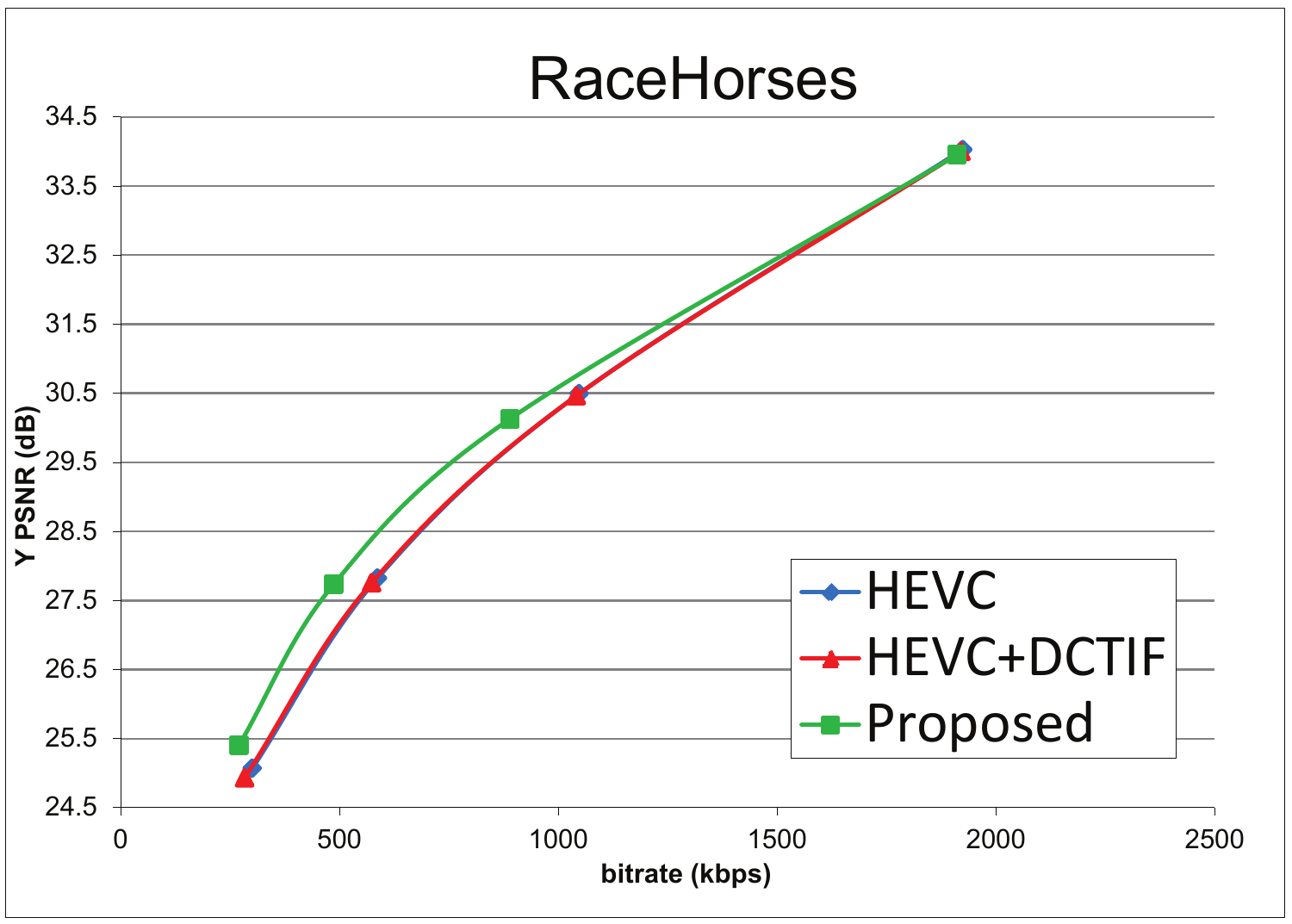}
}
\subfigure[ ]
{
\includegraphics[width=0.3\linewidth]{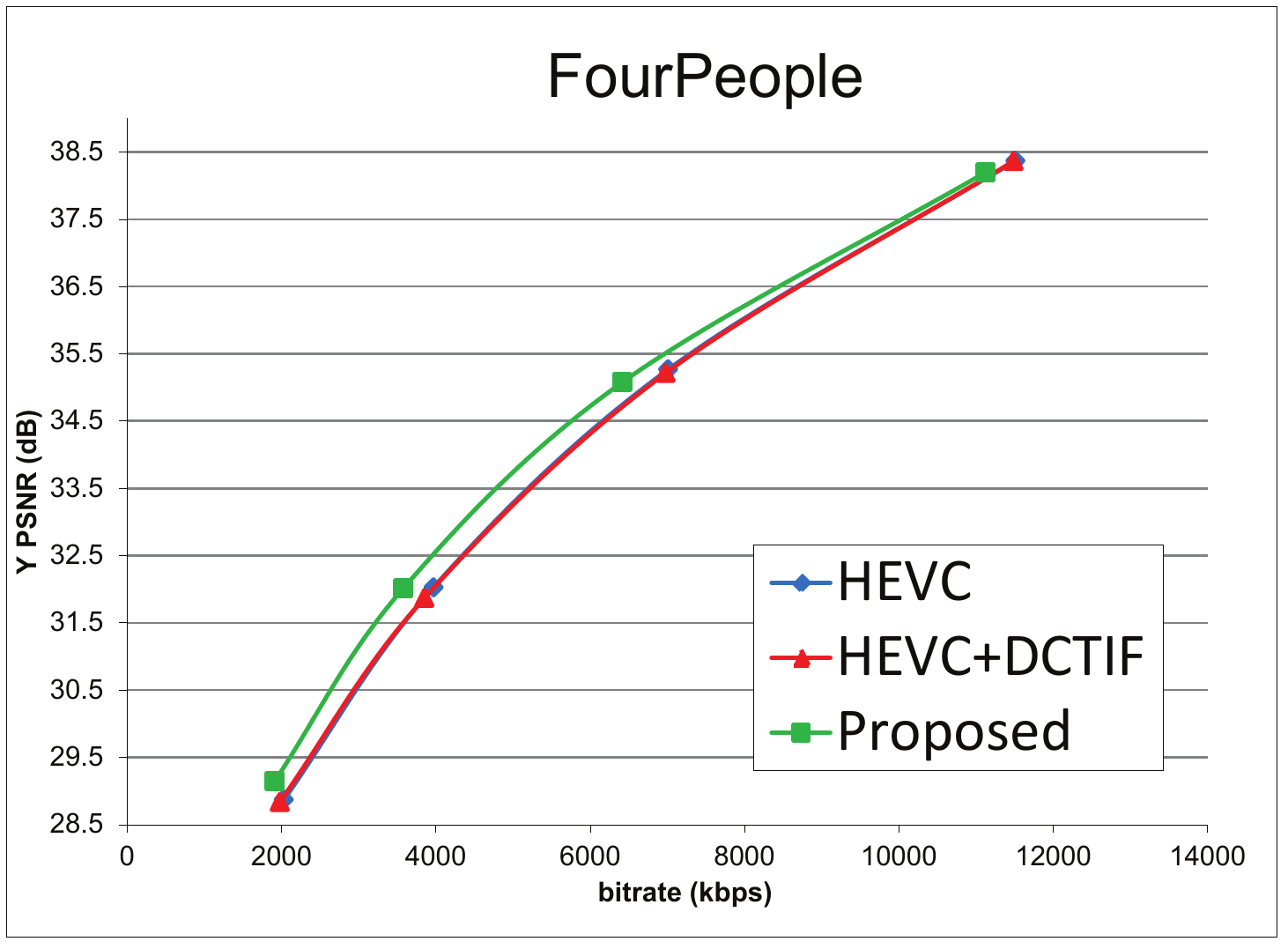}
}
\subfigure[ ]
{
\includegraphics[width=0.3\linewidth]{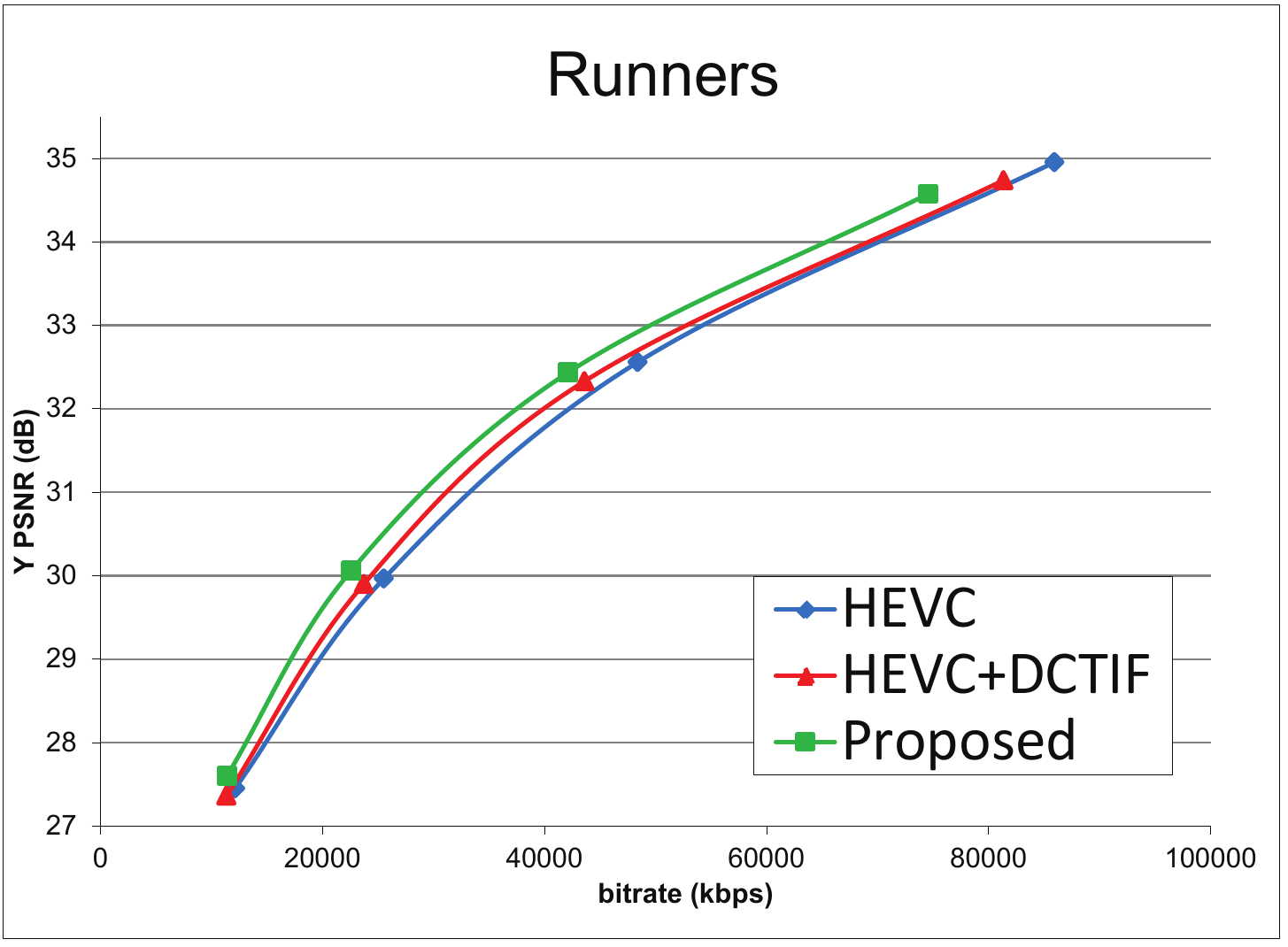}
}
\caption{Rate-distortion (R-D) curves of several typical sequences: (a) Traffic, (b) Kimono, (c) BasketballDrill, (d) RaceHorses, (e) FourPeople, and (f) Runners.}
\label{fig_rd}
\end{figure*}

Some typical R-D curves achieved by different schemes are shown in Fig. \ref{fig_rd}.
It can be observed that for most of the test sequences, our scheme achieves higher coding gain at lower bit rates, which is a nature of down/up-sampling-based coding. It is also visible that for different sequences, the switching bit-rates, at which the R-D curves of down/up-sampling-based coding and normal coding cross over, are quite diverse. Actually the switching bit-rate should be content dependent, which highlights the necessity of mode selection between low- and full-resolution coding.

In addition to the QPs adopted for the experiments in this paper (i.e. \{32, 37, 42, 47\}), we also tested the QPs 22 and 27 according to the HEVC common test conditions. Note that additional CNN models are trained for these two QPs. Table \ref{Results_of_Different_Bit_Rate_Ranges} summarizes the BD-rate results when comparing our scheme with HM anchor at different QPs. It can be observed, as QP increases, the BD-rate reductions become more and more significant. It again demonstrates down/up-sampling-based coding is useful especially at low bit rates.

\begin{table*}
\caption{BD-Rate Results of All Test Sequences}
\label{Results_of_All_Sequences}
\center
\begin{tabular}{c|l|rrrr|rrrr}
\hline
\multirow{2}{*}{Class}
         &  \multirow{2}{*}{Sequence}  &  \multicolumn{4}{c|}{BD-Rate (Anchored on HEVC)}                      &  \multicolumn{4}{c}{BD-Rate (Anchored on HEVC+DCTIF)}        \\
          \cline{3-10}
     &      &Y   &U         &V       & Y SSIM       &Y   &U         &V       & Y SSIM     \\
\hline
\multirow{4}{*}{Class A} & Traffic	          & --10.1\%   &--3.5\%	  &6.0\%	 &--12.9\%	                        & --8.0\%    &--13.2\%	  &--2.6\%	 &--7.9\%	\\
                         & PeopleOnStreet     &	--9.7\%    &--14.8\%  &--14.5\%  &--12.9\%                          & --8.5\%    &--20.4\%	  &--18.5\%	 &--9.7\%	\\
                         & Nebuta             &	--2.0\%    &--22.0\%  &3.1\%     &--4.4\%                           & --1.7\%    &--22.5\%	  &1.6\%	 &--3.6\%	\\
                         & SteamLocomotive    &	--1.7\%    &--27.7\%  &--25.4\%  &--6.1\%                           & --1.2\%    &--34.2\%	  &-25.6\%	 &--2.8\%	\\
\hline
\multirow{5}{*}{Class B} & Kimono             &	--7.7\%    &--5.5\%   &18.8\%    &--9.6\%                           & --3.4\%    &--25.9\%	  &--4.3\%	 &--3.4\%	\\
                         & ParkScene          &	--7.1\%    &--14.4\%  &--2.3\%   &--11.3\%                          & --5.0\%    &--25.2\%	  &--14.6\%	 &--6.6\%	\\
                         & Cactus             &	--6.6\%    &--2.5\%   &8.3\%     &--10.0\%                          & --5.0\%    &--6.5\%	  &0.9\%	 &--6.7\%	\\
                         & BQTerrace          &	--3.7\%    &--7.6\%   &--9.1\%   &--9.6\%                           & --3.1\%    &--8.2\%	  &--7.1\%	 &--6.5\%	\\
                         & BasketballDrive    &	--6.1\%    &--1.2\%   &3.2\%     &--10.8\%                          & --3.4\%    &--5.8\%	  &--2.5\%	 &--3.8\%	\\
 \hline
\multirow{4}{*}{Class C} & BasketballDrill    &	--4.9\%    &4.5\%     &8.1\%     &--7.9\%                           & --4.0\%   &4.9\%	      &2.1\%	 &--6.6\%	\\
                         & BQMall             &	--2.9\%    &--7.2\%   &--7.2\%   &--6.2\%                           & --2.3\%   &--10.6\%	  &--9.1\%	 &--5.3\%	\\
                         & PartyScene         &	--1.0\%    &--5.1\%   &--1.6\%   &--4.0\%                           & --1.0\%   &--5.5\%	  &--3.2\%	 &--3.6\%	\\
                         & RaceHorsesC        &	--6.7\%    &4.6\%     &7.5\%     &--10.7\%                          & --6.0\%   &1.9\%	      &3.9\%	 &--8.6\%	\\
\hline
\multirow{4}{*}{Class D} & BasketballPass     &	--2.0\%    &--3.7\%   &9.2\%     &--4.3\%                           & --2.3\%   &--7.5\%	  &12.3\%	 &--4.4\%	\\
                         & BQSquare           &	--0.9\%    &--0.6\%   &--21.1\%  &--1.4\%                           & --0.5\%   &1.7\%	      &-16.7\%	 &--1.2\%	\\
                         & BlowingBubbles     &	--3.2\%    &3.1\%     &--8.0\%   &--5.3\%                           & --1.7\%   &0.5\%	      &-9.6\%	 &--3.8\%	\\
                         & RaceHorses         &	--9.9\%    &7.5\%     &6.4\%     &--12.6\%                          & --9.6\%   &5.0\%	      &6.6\%	 &--11.1\%	\\
\hline
\multirow{3}{*}{Class E} & FourPeople         &	--7.2\%    &--10.5\%  &--11.0\%  &--11.0\%                          & --7.2\%   &--14.7\%	  &--14.5\%	 &--9.5\%	\\
                         & Johnny             &	--9.0\%    &--3.2\%   &--3.2\%   &--11.1\%                          & --7.1\%   &--6.0\%	  &--8.3\%	 &--5.6\%	\\
                         & KristenAndSara     &	--6.8\%    &--11.2\%  &--11.1\%  &--13.0\%                          & --5.3\%   &--8.4\%	  &--10.6\%	 &--8.2\%	\\
\hline
\multirow{5}{*}{Class UHD} & Fountains        &	--4.0\%    &--12.9\%  &--11.2\%  &--7.4\%                           & --2.0\%   &--16.1\%	  &--9.2\%	 &--2.0\%	\\
                           & Runners          &	--11.2\%   &22.8\%    &--0.1\%   &--12.4\%                          & --7.0\%   &0.9\%	      &--13.7\%	 &--6.0\%	\\
                           & Rushhour         &	--8.5\%    &4.4\%     &1.8\%     &--10.3\%                          & --3.2\%   &--9.2\%	  &--9.5\%	 &--3.0\%	\\
                           & TrafficFlow      &	--12.7\%   &--11.7\%  &--5.8\%   &--12.7\%                          & --6.9\%   &--17.3\%	  &--11.9\%	 &--5.6\%	\\
                           & CampfireParty    &	--8.4\%    &--10.8\%  &--0.8\%   &--9.5\%                           & --6.5\%   &--10.8\%	  &--5.0\%	 &--6.4\%	\\
\hline
   \multicolumn{2}{c|}{Average of Classes A--E}       & --5.5\%    &--6.0\%   &--2.2\%   &--8.8\%                           & --4.3\%   &--10.0\%	  &--6.0\%	 &--5.9\%	\\
\hline
   \multicolumn{2}{c|}{Average of Class UHD}        & --9.0\%    &--1.6\%   &--3.2\%   &--10.5\%                          & --5.1\%   &--10.5\%	  &--9.9\%	 &--4.6\%	\\
\hline
\end{tabular}
\end{table*}

\begin{table*}
\caption{BD-Rate Results at Different QPs (Anchored on HEVC)}
\label{Results_of_Different_Bit_Rate_Ranges}
\center
\begin{tabular}{c|rrrr|rrrr|rrrr}
\hline
\multirow{2}{*}{Class}
                         &  \multicolumn{4}{c|}{BD-Rate (QP 22--37)}         &  \multicolumn{4}{c}{BD-Rate (QP 27--42)}           &  \multicolumn{4}{c}{BD-Rate (QP 32--47)}        \\
          \cline{2-13}
                         &Y           &U          &V         & Y SSIM       &Y           &U           &V         & Y SSIM       &Y           &U           &V         & Y SSIM    \\
\hline
Class A                  & --0.4\%    &--3.3\%	  &--2.6\%	 &--1.8\%	    & --2.4\%    &--9.4\%	  &--5.5\%	 &--5.3\%	    & --5.9\%    &--17.0\%	  &--7.7\%	 &--9.1\%    \\
\hline
Class B                  & --1.4\%    &--3.3\%    &--0.7\%   &--2.8\%       & --3.5\%    &--5.0\%	  &0.6\%	 &--6.7\%	    & --6.2\%    &--6.2\%	  &3.8\%	 &--10.3\%    \\
 \hline
Class C                  & --0.2\%    &--0.5\%    &0.3\%     &--0.5\%       & --1.3\%    &--0.4\%     &1.6\%	 &--3.0\%	    & --3.9\%    &--0.8\%	  &1.7\%	 &--7.2\%    \\
\hline
Class D                  & --0.3\%    &0.3\%      &--0.9\%   &--1.0\%       & --1.4\%    &1.0\%	      &--2.3\%	 &--3.7\%	    & --4.0\%    &1.6\%	      &--3.4\%	 &--6.4\%    \\
\hline
Class E                  & --1.0\%    &--3.3\%    &--4.9\%   &--2.7\%       & --3.8\%    &--6.0\%	  &--8.2\%	 &--7.6\%	    & --7.7\%    &--8.3\%	  &--8.4\%	 &--11.7\%    \\
\hline
Avg. Classes A--E                 & --0.7\%    &--2.0\%    &--1.6\%   &--1.7\%       & --2.5\%    &--3.9\%	  &--2.3\%	 &--5.0\%	    & --5.5\%    &--6.0\%	  &--2.2\%	 &--8.8\%    \\
\hline
Class UHD                & --2.1\%    &--6.6\%    &--4.5\%   &--4.0\%       & --5.6\%    &--6.8\%	  &--4.9\%	 &--7.8\%	    & --9.0\%    &--1.6\%	  &--3.2\%	 &--10.5\%    \\
\hline
\end{tabular}
\end{table*}

\subsubsection{Mode Selection Results}
\label{sec_Mode_Hitting}
Since our proposed scheme decides whether to down-sample at block level, we perform analyses of the blocks that choose low-resolution coding mode to further understand the performance. Some symbols are defined as shown in Table \ref{CTU_Category}, and the hitting ratios are calculated as follows,
\begin{equation*}
\label{eqn_hitting_luma}
P_{Hitting} = \frac{\#C_{Hitting}}{\#C_{Total}},
P_{Luma} = \frac{\#C_{Luma}}{\#C_{Hitting}},
\end{equation*}
\begin{equation*}
\label{eqn_cb_cr}
P_{Cb} = \frac{\#C_{Cb}}{\#C_{Hitting}},
P_{Cr} = \frac{\#C_{Cr}}{\#C_{Hitting}}
\end{equation*}
where the symbol $\#$ denotes counting the amount.
Table \ref{Results_of_mode_hitting} presents the calculated hitting ratios.
$P_{Hitting}$ is on average 72.2\%, 68.4\%, 48.1\%, 42.4\%, 68.7\%, 85.2\% for Classes A, B, C, D, E, UHD, respectively.
Taken into account the resolutions of these videos, it is obvious that the hitting ratio becomes higher as the video resolution increases. It shows the effectiveness of down/up-sampling-based coding for high definition content, and also interprets the reason that our scheme achieves higher BD-rate reduction on UHD sequences.
Moreover, among the blocks choosing low-resolution coding, a majority of them choose CNN-based up-sampling method, as can be observed from the last three columns of Table \ref{Results_of_mode_hitting}.
Meanwhile, DCTIF is also useful for certain video content and especially for chroma components.

\begin{figure*}
\centering
\subfigure[ ]
{
\includegraphics[width=0.8\linewidth]{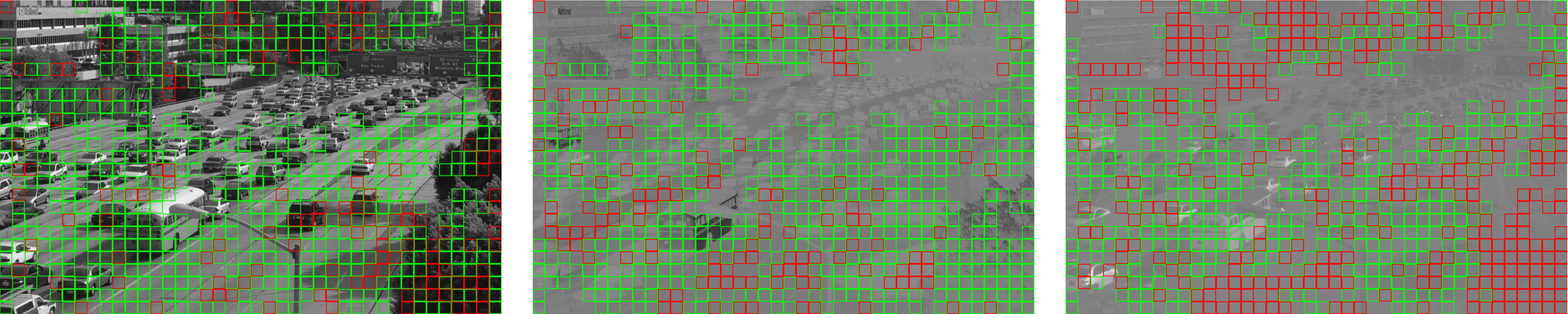}
}
\subfigure[ ]
{
\includegraphics[width=0.8\linewidth]{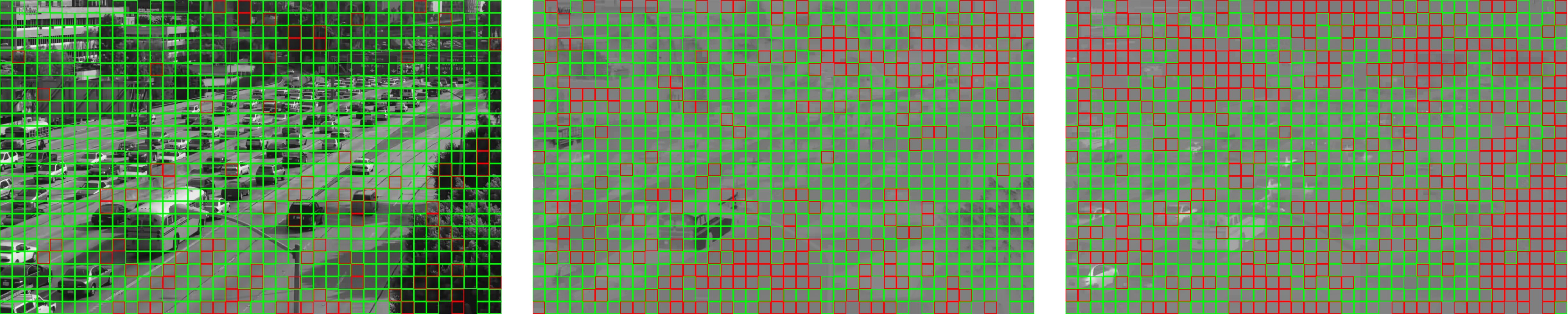}
}
\subfigure[ ]
{
\includegraphics[width=0.8\linewidth]{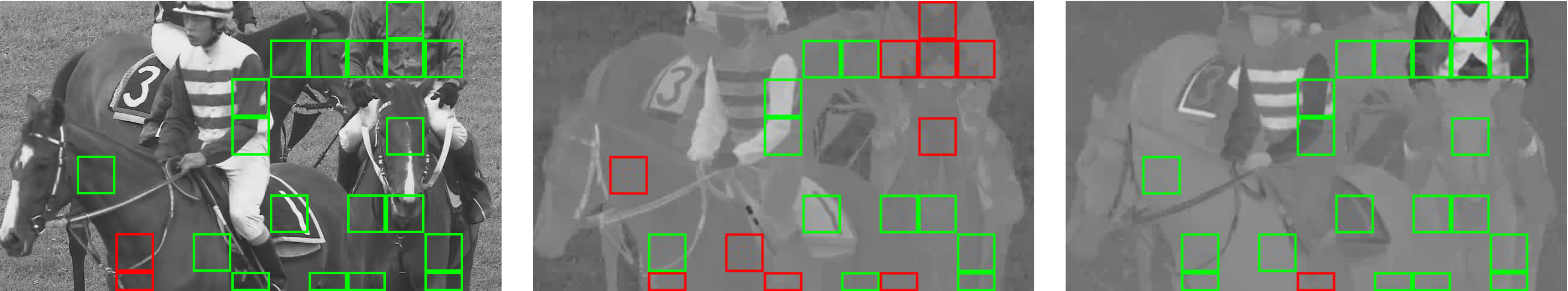}
}
\subfigure[ ]
{
\includegraphics[width=0.8\linewidth]{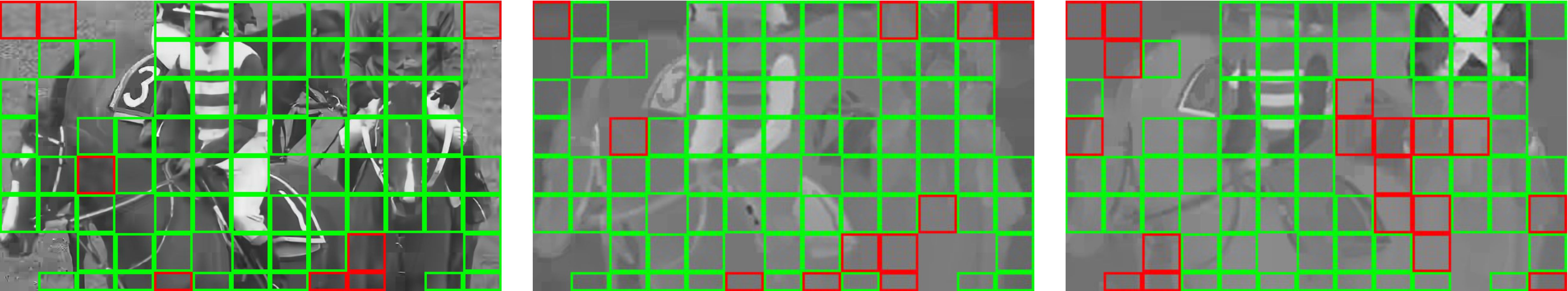}
}
\caption{This figure shows the CTUs that choose different modes. CTUs with green block are coded at low resolution and up-sampled using CNN, CTUs with red block are also coded at low resolution but up-sampled using DCTIF, and other CTUs are coded at full resolution. From left to right: Y, Cb, and Cr. Cb and Cr are shown in the same size as Y for display purpose only. From top to bottom:
(a) Traffic,     QP = 32, $P_{Hitting}$ = 64.6\%, $P_{Luma}$ = 80.5\%, $P_{Cb}$ = 79.9\%, $P_{Cr}$ = 55.0\%,
(b) Traffic,     QP = 42, $P_{Hitting}$ = 95.2\%, $P_{Luma}$ = 90.3\%, $P_{Cb}$ = 76.2\%, $P_{Cr}$ = 58.9\%,
(c) RaceHorsesC, QP = 32, $P_{Hitting}$ = 20.2\%, $P_{Luma}$ = 90.5\%, $P_{Cb}$ = 52.4\%, $P_{Cr}$ = 95.2\%,
(d) RaceHorsesC, QP = 42, $P_{Hitting}$ = 79.8\%, $P_{Luma}$ = 90.4\%, $P_{Cb}$ = 86.7\%, $P_{Cr}$ = 78.3\%.
}
\label{fig_mode_hitting}
\end{figure*}

Fig. \ref{fig_mode_hitting} is provided for visually inspecting the blocks that choose different coding modes and different up-sampling methods.
We can observe that CNN-based method is good at reconstructing structural regions, whereas DCTIF is prone to be selected for smooth and some textural regions.
For example, in Fig. \ref{fig_mode_hitting} (a), most of the CTUs containing vehicles choose CNN-based up-sampling, while most of the CTUs corresponding to road choose DCTIF.
Due to different properties of the luma and chroma components, the selections of up-sampling methods are not always consistent among Y, Cb, and Cr.
Note the bottom right corner in Fig. \ref{fig_mode_hitting} (a) and (b), the CTUs mostly choose CNN-based up-sampling for Y and Cb, but choose DCTIF for Cr, since the Cr component of these CTUs is quite smooth.
In addition, low-resolution coding becomes more competitive when the bit rate is lower, as can be observed by comparing the hitting ratios in Fig. \ref{fig_mode_hitting} (a) versus (b), and (c) versus (d).

\begin{table}
\centering
\caption{Symbols for CTUs that Choose Different Modes}
\begin{tabular}{c|l}
\hline
Symbol                  & Remark \\
\hline
$C_{Total}$                        & All CTUs in a frame \\
\hline
$C_{Hitting}$     & CTUs selecting the mode of low-resolution coding \\
\hline
\multirow{2}{*}{$C_{Luma}$}        & Low-resolution coded CTUs, whose luma component\\
                                   &is up-sampled using CNN\\
\hline
\multirow{2}{*}{$C_{Cb}$}          & Low-resolution coded CTUs, whose Cb component\\
                                   &is up-sampled using CNN\\
\hline
\multirow{2}{*}{$C_{Cr}$}          & Low-resolution coded CTUs, whose Cr component\\
                                   &is up-sampled using CNN\\
\hline
\end{tabular}
\label{CTU_Category}
\end{table}

\begin{table}
\caption{Hitting Ratio Results on Different Classes of Test Sequences}
\label{Results_of_mode_hitting}
\center
\begin{tabular}{c|l|c|c|c}
\hline
Class          & $P_{Hitting}$                  & $P_{Luma}$          & $P_{Cb}$           & $P_{Cr}$         \\
\hline
Class A        & 72.2\%                    &70.3\%	           &71.2\%	          &55.0\%	   \\
\hline
Class B        & 68.4\%                    &75.0\%	           &65.1\%	          &49.4\%	   \\
\hline
Class C        & 48.1\%                    &92.0\%	           &68.5\%	          &73.5\%	   \\
\hline
Class D        & 42.4\%                    &81.9\%	           &51.6\%	          &70.7\%	   \\
\hline
Class E        & 68.7\%                    &72.8\%	           &54.4\%	          &58.5\%	   \\
\hline
Class UHD      & 85.2\%                    &68.4\%	           &54.2\%	          &64.1\%	   \\
\hline
\end{tabular}
\end{table}
\subsubsection{Generalization of CNN for Different QPs}
\label{sec_robustness_of_training_model}
We have trained different CNN models for different QPs in the above experiments. In practice, it may be too costly to train a different model for every QP. Thus, we investigate the generalization ability of CNN for different QPs. In the following experiments, we use the models trained at four QPs: \{32, 37, 42, 47\}, but the QPs during compression are set to \{34, 39, 44, 49\} (denoted by QP$+2$), or \{30, 35, 40, 45\} (denoted by QP$-2$). For each test QP, the models trained at the nearest QP are retrieved for usage.
Table \ref{Results_Robustness} summarizes the experimental results.
BD-rate reductions are still observed from these results, showing the effectiveness of the trained models when used for different QPs. Therefore, the amount of models required in practice can be much less than the number of possible QPs. Furthermore, the BD-rate reductions of QP$+2$ are usually more significant than those of QP$-2$, since higher QP corresponds to lower bit rate that prefers low-resolution coding.
\begin{table}
\caption{BD-Rate Results of Using Trained CNN Models for Different QPs}
\label{Results_Robustness}
\center
\begin{tabular}{c|cc|cc}
\hline
\multirow{2}{*}{Class}   &  \multicolumn{2}{c|}{Anchored on HEVC} &  \multicolumn{2}{c}{Anchored on HEVC+DCTIF}    \\
          \cline{2-5}
                         &QP$+2$         &QP$-2$         &QP$+2$         &QP$-2$                                             \\
\hline
Class A                  &--6.6\%      &--4.5\%	     &--5.2\%	   &--3.8\%	                        	             \\
\hline
Class B                  &--6.9\%      &--5.5\%      &--4.0\%      &--3.3\%                                          \\
 \hline
Class C                  &--5.5\%      &--2.9\%      &--4.9\%      &--2.9\%                            	             \\
\hline
Class D                  &--6.0\%      &--2.1\%      &--5.0\%      &--2.0\%                                          \\
\hline
Class E                  &--8.1\%      &--6.5\%      &--6.7\%      &--5.6\%                                          \\
\hline
Avg. Classes A--E                 &--6.6\%      &--4.3\%      &--5.0\%      &--3.4\%                           	             \\
\hline
Class UHD                &--9.0\%      &--8.5\%      &--4.9\%      &--5.0\%                           	            \\
\hline
\end{tabular}
\end{table}
\subsubsection{Verification of the Designed CNN}
\label{sec_verify_cnn}
In order to verify the performance of our designed CNN, we have compared it with the fixed interpolation filter DCTIF as well as a state-of-the-art CNN-based image SR method, i.e. VDSR \cite{kim2016accurate}. VDSR is a deep network consisting of 20 layers and is shown to outperform the shallow network, SRCNN \cite{dong2014learning}, by a large margin. For fair comparison, we follow the instructions in \cite{kim2016accurate} to train VDSR, but using our own training data produced when QP is 32. The comparative experiments are performed as follows. The test sequences are entirely down-sampled and then compressed with QP equal to 32, and then up-sampled by each method. The comparative results of the luma component are summarized in Table \ref{Results_Comparison_VDSR}. It can be observed that both VDSR and our CNN-based method outperform DCTIF significantly. Our CNN-based method is better than VDSR for most of the test sequences, and achieves on average 0.16 dB gain. It is worth noting that our network is shallower and simpler than VDSR, but is very competitive due to the adopted multi-scale fusion and deconvolution, which are not used in VDSR.

We have also verified our designed chroma up-sampling CNN experimentally. In previous work on image SR, the chroma components are usually up-sampled by fixed interpolation filters. So we compare three methods: DCTIF, CNN without luma, and CNN with luma. The CNN without luma method has a similar network structure to that shown in Fig.~\ref{fig_chroma_network} but excluding the luma information from the network input. The CNN without luma network is also trained under the same setting and using the same training data. The experimental settings are identical to those in the previous paragraph, and comparative results are shown in Table \ref{Results_of_incorporating}.
It can be observed that CNN-based methods outperform DCTIF consistently, but the PSNR gain is not as much as that for luma (in Table \ref{Results_Comparison_VDSR}), since the chroma components of natural images are usually quite smooth and the potential improvement is limited. Moreover, the proposed CNN using luma achieves better performance than the CNN without luma, leading to on average 0.20 dB and 0.22 dB gain for Cb and Cr, respectively.
Such results confirm the effectiveness of using luma information to boost the chroma up-sampling performance.

\begin{table}
\caption{PSNR Results of Different Up-sampling Methods for Luma}
\label{Results_Comparison_VDSR}
\center
\begin{tabular}{c|l|c|c|c}
\hline
Class   &Sequence    &  DCTIF      &  VDSR  &  Ours    \\
\hline
\multirow{5}{*}{Class B} & Kimono	                   &39.82      &\textbf{39.86}       &39.79       \\
                         & ParkScene                   &34.23      &34.26                &\textbf{34.63}       \\
                         & Cactus                      &33.01      &33.59                &\textbf{33.86}       \\
                         & BQTerrace                   &28.50      &29.61                &\textbf{30.20}       \\
                         & BasketballDrive             &34.36      &35.64                &\textbf{35.71}       \\
\hline
\multirow{4}{*}{Class C} & BasketballDrill	           &31.46      &32.77                &\textbf{33.17}      \\
                         & BQMall                      &28.31      &29.36                &\textbf{29.46}      \\
                         & PartyScene                  &24.94      &26.00                &\textbf{26.16}      \\
                         & RaceHorsesC                 &30.06      &31.09                &\textbf{31.18}      \\
\hline
\multirow{4}{*}{Class D} & BasketballPass	           &30.34      &31.33                &\textbf{31.56}      \\
                         & BQSquare                    &23.48      &\textbf{25.95}       &25.80      \\
                         & BlowingBubbles              &28.52      &29.50                &\textbf{29.57}      \\
                         & RaceHorses                  &29.59      &31.34                &\textbf{31.56}      \\
\hline
\multirow{3}{*}{Class E} & FourPeople	               &36.45      &37.58                &\textbf{37.74}      \\
                         & Johnny                      &34.30      &\textbf{36.90}       &36.64      \\
                         & KristenAndSara              &34.12      &35.43                &\textbf{35.69}      \\
\hline
   \multicolumn{2}{c|}{Average}                        &31.34      &32.51                &\textbf{32.67}      \\
\hline
\end{tabular}
\end{table}

\begin{table*}
\caption{PSNR Results of Different Up-sampling Methods for Chroma}
\label{Results_of_incorporating}
\center
\begin{tabular}{c|l|cc|cc|cc}
\hline
\multirow{2}{*}{Class}   &\multirow{2}{*}{Sequence}    &  \multicolumn{2}{c|}{DCTIF}     &  \multicolumn{2}{c|}{CNN without luma} &  \multicolumn{2}{c}{CNN with luma}    \\
          \cline{3-8}
                         &                             &Cb            &Cr                &Cb           &Cr                     &Cb           &Cr                 \\
\hline
\multirow{5}{*}{Class B} & Kimono	                   &40.60      &41.75                &40.60      &41.70                    &\textbf{41.13}      &\textbf{41.86}             \\
                         & ParkScene                   &38.20      &39.74                &38.20      &39.68                    &\textbf{38.83}      &\textbf{39.80} \\
                         & Cactus                      &37.68      &39.32                &37.77      &39.51                    &\textbf{37.83}      &\textbf{39.58} \\
                         & BQTerrace                   &38.19      &40.61                &38.22      &40.73                    &\textbf{38.50}      &\textbf{40.85} \\
                         & BasketballDrive             &41.86      &41.71                &\textbf{42.00}      &42.02           &41.90               &\textbf{42.23} \\
\hline
\multirow{4}{*}{Class C} & BasketballDrill	           &37.41      &37.51                &\textbf{37.63}      &37.79           &37.49      &\textbf{38.34}\\
                         & BQMall                      &38.36      &39.34                &38.66      &39.79                    &\textbf{39.08}      &\textbf{40.16}\\
                         & PartyScene                  &34.89      &35.45                &35.03      &35.65                    &\textbf{35.34}      &\textbf{35.91}\\
                         & RaceHorsesC                 &36.58      &37.66                &36.80      &38.22                    &\textbf{37.00}      &\textbf{38.40}\\
\hline
\multirow{4}{*}{Class D} & BasketballPass	           &37.72      &36.94                &37.88      &37.41                    &\textbf{38.02}      &\textbf{37.80}\\
                         & BQSquare                    &39.22      &39.53                &39.33      &39.84                    &\textbf{39.69}      &\textbf{40.16}\\
                         & BlowingBubbles              &35.26      &37.32                &35.36      &37.52                    &\textbf{35.38}      &\textbf{37.73}\\
                         & RaceHorses                  &35.53      &35.38                &35.82      &35.99                    &\textbf{36.39}      &\textbf{36.30}\\
\hline
\multirow{3}{*}{Class E} & FourPeople	               &43.86      &44.65                &\textbf{44.12}      &44.95           &43.96               &\textbf{45.01}\\
                         & Johnny                      &43.14      &44.22                &\textbf{43.33}      &44.51           &43.22      &\textbf{44.55}\\
                         & KristenAndSara              &42.19      &43.51                &42.43      &43.84                    &\textbf{42.62}      &\textbf{44.04}\\
\hline
   \multicolumn{2}{c|}{Average}                        &38.79      &39.67                &38.95      &39.95                    &\textbf{39.15}      &\textbf{40.17}        	\\
\hline
\end{tabular}
\end{table*}
\subsubsection{Verification of Two-Stage Up-sampling}
\label{sec_effectiveness_of_refining}
We have verified the proposed two-stage up-sampling strategy by comparing with only one stage of up-sampling. Table \ref{Percent_of_refining} presents the average MSE of the reconstructed CTUs that choose low-resolution coding mode, after the first stage and after the second stage, respectively. The percentage of CTUs that benefit from the second stage (i.e. MSE decreases) is also shown in the table.
Table \ref{Results_of_refining} further presents the BD-rate results of using the second stage.
The BD-rate reductions provided by the second stage of up-sampling are on average 0.7\%, 2.7\%, 3.0\% for HEVC test sequences, and 0.8\%, 3.4\%, 3.7\% for UHD test sequences, on Y, U, and V, respectively.
As shown, the BD-rate reductions on chroma components are higher than luma. This is due to the lower resolution of chroma (32$\times$32 for CTU) that incurs more severe influence by the lack of boundary information.
Note that in our current implementation, the result of the first stage is simply replaced by that of the second stage. But as can be observed in Table \ref{Percent_of_refining}, there are a portion of blocks for which the second stage incurs worse result. We may adaptively decide whether to perform the second stage for each block, which will be studied in the future.
\begin{table}
\centering
\caption{Percentage of CTUs that Benefit from the Second Stage, and Average MSE Results (Luma, QP 37)}
\label{Percent_of_refining}
\center
\begin{tabular}{c|l|l|l}
\hline
\multirow{2}{*}{Class} & \multirow{2}{*}{Percentage} & Average MSE & Average MSE \\
                       &  & (First Stage)           & (Second Stage)                           \\
\hline
Class A  	             & 72.8\%                    &39.95	           &\textbf{39.73}	                    \\
\hline
Class B  	             & 72.1\%                    &25.01	           &\textbf{24.80}	                    \\
\hline
Class C  	             & 85.8\%                    &36.04	           &\textbf{34.70}	                    \\
\hline
Class D  	             & 76.9\%                    &40.38	           &\textbf{39.76}	                    \\
\hline
Class E  	             & 76.7\%                    &15.17	           &\textbf{14.84}	                    \\
\hline
Class UHD  	             & 70.2\%                    &22.57	           &\textbf{22.42}	                    \\
\hline
Average                  & 75.3\%                    &29.96            &\textbf{29.49}                    	\\
\hline
\end{tabular}
\end{table}
\begin{table}
\caption{BD-Rate Results of Using The Second Stage of Up-sampling}
\label{Results_of_refining}
\center
\begin{tabular}{c|l|rrr}
\hline
Class                    &   Sequence        &Y           &U         &V                                    \\
\hline
\multirow{4}{*}{Class A} & Traffic	          & --0.9\%    &--3.3\%	  &--2.7\%	                     	    \\
                         & PeopleOnStreet     &	--1.0\%    &--5.4\%   &--5.2\%                          	\\
                         & Nebuta             &	--0.3\%    &--4.0\%   &--1.0\%                            	\\
                         & SteamLocomotive    &	--0.4\%    &--4.4\%   &--7.3\%                            	\\
\hline
\multirow{5}{*}{Class B} & Kimono             &	--0.7\%    &--5.6\%   &--2.4\%                              \\
                         & ParkScene          &	--0.7\%    &--4.4\%   &--2.6\%                              \\
                         & Cactus             &	--0.7\%    &--2.4\%   &--2.7\%                              \\
                         & BQTerrace          &	--0.6\%    &--1.9\%   &--2.5\%                              \\
                         & BasketballDrive    &	--0.7\%    &--1.5\%   &--2.2\%                             	\\
 \hline
\multirow{4}{*}{Class C} & BasketballDrill    &	--0.9\%    &--1.1\%   &--2.2\%                             	\\
                         & BQMall             &	--0.6\%    &--2.4\%   &--3.6\%                             	\\
                         & PartyScene         &	--0.2\%    &--1.0\%   &--1.0\%                             	\\
                         & RaceHorsesC        &	--1.0\%    &--3.7\%   &--4.3\%                              \\
\hline
\multirow{4}{*}{Class D} & BasketballPass     &	--0.6\%    &0.2\%     &--1.2\%                              \\
                         & BQSquare           &	--0.2\%    &--1.5\%   &--2.7\%                            	\\
                         & BlowingBubbles     &	--0.5\%    &--0.1\%   &--3.1\%                             	\\
                         & RaceHorses         &	--1.2\%    &--2.8\%   &--2.5\%                             	\\
\hline
\multirow{3}{*}{Class E} & FourPeople         &	--1.2\%    &--3.0\%   &--4.3\%                           	\\
                         & Johnny             &	--1.1\%    &--2.9\%   &--2.9\%                              \\
                         & KristenAndSara     &	--0.9\%    &--3.3\%   &--3.5\%                           	\\
\hline
\multirow{5}{*}{Class UHD} & Fountains        &	--0.6\%    &--5.1\%   &--4.3\%                              \\
                           & Runners          &	--0.8\%    &--2.2\%   &--3.4\%                              \\
                           & Rushhour         &	--0.7\%    &--3.2\%   &--3.7\%                              \\
                           & TrafficFlow      &	--1.1\%    &--4.6\%   &--4.3\%                              \\
                           & CampfireParty    &	--0.7\%    &--1.8\%   &--3.0\%                              \\
\hline
   \multicolumn{2}{c|}{Average of Classes A--E}       & --0.7\%    &--2.7\%   &--3.0\%                             	\\
\hline
   \multicolumn{2}{c|}{Average of Class UHD}        & --0.8\%    &--3.4\%   &--3.7\%                            	\\
\hline
\end{tabular}
\end{table}

\subsubsection{Computational Complexity}
\label{sec_complexity}
One drawback of CNN-based up-sampling methods is the high computational complexity compared to simple interpolation filters such as DCTIF. In our current implementation, the CNN is not optimized for computational speed, and thus the encoding/decoding time of our scheme is much longer than that of the highly optimized HEVC anchor. The computational time comparison is summarized in Table \ref{Results_CNN_Complexity}. It can be observed the increase of encoding time varies little across different videos, but the increase of decoding time varies much. In fact, most of the decoding computations are cost on the CNN up-sampling, and thus the increase of decoding time is dependent on the amount of blocks that choose CNN up-sampling.

When designing our CNN, we have tried to keep it simple while pursuing high reconstruction quality. Accordingly, we have compared the computational complexity of our CNN and VDSR, since both networks achieve comparable reconstruction quality as presented in Section \ref{sec_verify_cnn}. Experimental results show that, it takes on average 0.032 and 0.123 seconds for our CNN and VDSR, respectively, to up-sample the luma component of a down-sampled and coded CTU, on a workstation with an Intel CPU at 4.0 GHz. Our CNN is almost 4 times faster than VDSR, the number being proportional to the amount of layers (our CNN has 5 layers and VDSR has 20 layers). It again demonstrates the advantage of our designed CNN.
\begin{table}
\centering
\caption{Computational Time Comparison of Our Scheme and HEVC (HEVC is 100\%)}
\label{Results_CNN_Complexity}
\begin{tabular}{c|l|l}
\hline
Class                    & Encoding Time       & Decoding Time                   \\
\hline
Class A                  &713\%                       &29572\%	                           	 \\
\hline
Class B                  &749\%                       &29868\%                                 \\
 \hline
Class C                  &726\%                       &16754\%                                 \\
\hline
Class D                  &751\%                       &13721\%                                 \\
\hline
Class E                  &779\%                       &21502\%                                 \\
\hline
Class UHD                &766\%                       &44766\%                    	         \\
\hline
Average                  &747\%                       &25021\%                    	         \\
\hline
\end{tabular}
\end{table}
\section{Conclusion}
\label{sec_Conclusion}

In this paper, we propose a CNN-based block up-sampling scheme for intra frame coding.
A CTU is optionally down-sampled before being compressed by normal intra coding, and then up-sampled to its original resolution, so as to enable adaptive sampling rates for different CTUs.
We carefully design the CNN structures for the up-sampling of luma and chroma, and propose to combine the CNN-based up-sampling with the fixed interpolation filter (DCTIF) to allow for adaptive up-sampling methods for the down-sampled CTUs.
We also propose a two-stage up-sampling process in accordance to the block-level down/up-sampling, and study the coding parameters setting of the down-sampled CTUs for pursing frame-level R-D optimization.
Our proposed methods are implemented based on HEVC reference software, and extensive experimental results demonstrate the superior performance achieved by the proposed methods than HEVC anchor. On average 5.5\% and 9.0\% BD-rate reductions are achieved by our scheme for HEVC common test sequences and UHD test sequences, respectively. The proposed scheme is especially useful for compressing high definition videos at low bit rates.

In the future, we plan to extend this work in two directions. First, the idea of CNN-based block-level down/up-sampling may be extended for inter frame coding in addition to intra; how to integrate down/up-sampling with motion estimation/compensation is a key issue therein. Second, though the proposed CNN structure in this paper is much simpler than the state-of-the-arts, its computational complexity is still quite high; how to accelerate the speed of CNN-based up-sampling needs to be addressed. We anticipate that new hardware/firmware based implementation of CNN is probably the solution.
\ifCLASSOPTIONcaptionsoff
  \newpage
\fi


\begin{thebibliography}{10}
\providecommand{\url}[1]{#1}
\csname url@samestyle\endcsname
\providecommand{\newblock}{\relax}
\providecommand{\bibinfo}[2]{#2}
\providecommand{\BIBentrySTDinterwordspacing}{\spaceskip=0pt\relax}
\providecommand{\BIBentryALTinterwordstretchfactor}{4}
\providecommand{\BIBentryALTinterwordspacing}{\spaceskip=\fontdimen2\font plus
\BIBentryALTinterwordstretchfactor\fontdimen3\font minus
  \fontdimen4\font\relax}
\providecommand{\BIBforeignlanguage}[2]{{%
\expandafter\ifx\csname l@#1\endcsname\relax
\typeout{** WARNING: IEEEtran.bst: No hyphenation pattern has been}%
\typeout{** loaded for the language `#1'. Using the pattern for}%
\typeout{** the default language instead.}%
\else
\language=\csname l@#1\endcsname
\fi
#2}}
\providecommand{\BIBdecl}{\relax}
\BIBdecl

\bibitem{sugawara2014ultra}
M.~Sugawara, S.-Y. Choi, and D.~Wood, ``Ultra-high-definition television ({Rec.
  ITU-R BT. 2020}): A generational leap in the evolution of television,''
  \emph{IEEE Signal Processing Magazine}, vol.~31, no.~3, pp. 170--174, 2014.

\bibitem{sullivan2012overview}
G.~J. Sullivan, J.~Ohm, W.-J. Han, and T.~Wiegand, ``Overview of the high
  efficiency video coding ({HEVC}) standard,'' \emph{IEEE Transactions on
  Circuits and Systems for Video Technology}, vol.~22, no.~12, pp. 1649--1668,
  2012.

\bibitem{bruckstein2003down}
A.~M. Bruckstein, M.~Elad, and R.~Kimmel, ``Down-scaling for better transform
  compression,'' \emph{IEEE Transactions on Image Processing}, vol.~12, no.~9,
  pp. 1132--1144, 2003.

\bibitem{takahashi2011rate}
K.~Takahashi, T.~Naemura, and M.~Tanaka, ``Rate-distortion analysis of
  super-resolution image/video decoding,'' in \emph{IEEE International
  Conference on Image Processing}.\hskip 1em plus 0.5em minus 0.4em\relax IEEE,
  2011, pp. 1629--1632.

\bibitem{molina2006toward}
R.~Molina, A.~K. Katsaggelos, L.~D. Alvarez, and J.~Mateos, ``Toward a new
  video compression scheme using super-resolution,'' in \emph{Proceedings of
  SPIE}, vol. 6077, 2006, pp. 67--79.

\bibitem{barreto2007region}
D.~Barreto, L.~D. Alvarez, R.~Molina, A.~K. Katsaggelos, and G.~M. Callico,
  ``Region-based super-resolution for compression,'' \emph{Multidimensional
  Systems and Signal Processing}, vol.~18, no. 2-3, pp. 59--81, 2007.

\bibitem{shen2011down}
M.~Shen, P.~Xue, and C.~Wang, ``Down-sampling based video coding using
  super-resolution technique,'' \emph{IEEE Transactions on Circuits and Systems
  for Video Technology}, vol.~21, no.~6, pp. 755--765, 2011.

\bibitem{wu2009image}
J.~Wu, Y.~Xing, G.~Shi, and L.~Jiao, ``Image compression with downsampling and
  overlapped transform at low bit rates,'' in \emph{IEEE International
  Conference on Image Processing}.\hskip 1em plus 0.5em minus 0.4em\relax IEEE,
  2009, pp. 29--32.

\bibitem{lin2006adaptive}
W.~Lin and L.~Dong, ``Adaptive downsampling to improve image compression at low
  bit rates,'' \emph{IEEE Transactions on Image Processing}, vol.~15, no.~9,
  pp. 2513--2521, 2006.

\bibitem{nguyen2008adaptive}
V.-A. Nguyen, Y.-P. Tan, and W.~Lin, ``Adaptive downsampling/upsampling for
  better video compression at low bit rate,'' in \emph{IEEE International
  Symposium on Circuits and Systems}.\hskip 1em plus 0.5em minus 0.4em\relax
  IEEE, 2008, pp. 1624--1627.

\bibitem{yang2010image}
J.~Yang, J.~Wright, T.~S. Huang, and Y.~Ma, ``Image super-resolution via sparse
  representation,'' \emph{IEEE Transactions on Image Processing}, vol.~19,
  no.~11, pp. 2861--2873, 2010.

\bibitem{dong2014learning}
C.~Dong, C.~C. Loy, K.~He, and X.~Tang, ``Learning a deep convolutional network
  for image super-resolution,'' in \emph{European Conference on Computer
  Vision}.\hskip 1em plus 0.5em minus 0.4em\relax Springer, 2014, pp. 184--199.

\bibitem{dong2016accelerating}
C.~Dong, C.~C. Loy, and X.~Tang, ``Accelerating the super-resolution
  convolutional neural network,'' in \emph{European Conference on Computer
  Vision}.\hskip 1em plus 0.5em minus 0.4em\relax Springer, 2016, pp. 391--407.

\bibitem{wang2016end}
Y.~Wang, L.~Wang, H.~Wang, and P.~Li, ``End-to-end image super-resolution via
  deep and shallow convolutional networks,'' \emph{arXiv preprint
  arXiv:1607.07680}, 2016.

\bibitem{kim2016accurate}
J.~Kim, J.~K. Lee, and K.~M. Lee, ``Accurate image super-resolution using very
  deep convolutional networks,'' in \emph{IEEE Conference on Computer Vision
  and Pattern Recognition}, 2016, pp. 1646--1654.

\bibitem{kim2016deeply}
------, ``Deeply-recursive convolutional network for image super-resolution,''
  in \emph{IEEE Conference on Computer Vision and Pattern Recognition}, 2016,
  pp. 1637--1645.

\bibitem{nair2010rectified}
V.~Nair and G.~E. Hinton, ``Rectified linear units improve restricted boltzmann
  machines,'' in \emph{International Conference on Machine Learning}, 2010, pp.
  807--814.

\bibitem{wang2004image}
Z.~Wang, A.~C. Bovik, H.~R. Sheikh, and E.~P. Simoncelli, ``Image quality
  assessment: From error visibility to structural similarity,'' \emph{IEEE
  Transactions on Image Processing}, vol.~13, no.~4, pp. 600--612, 2004.

\bibitem{schaefer2004ucid}
G.~Schaefer and M.~Stich, ``{UCID}: An uncompressed color image database,'' in
  \emph{Proceedings of SPIE}, vol. 5307, 2004, pp. 472--480.

\bibitem{segall2004bayesian}
C.~A. Segall, A.~K. Katsaggelos, R.~Molina, and J.~Mateos, ``Bayesian
  resolution enhancement of compressed video,'' \emph{IEEE Transactions on
  Image Processing}, vol.~13, no.~7, pp. 898--911, 2004.

\bibitem{he2016deep}
K.~He, X.~Zhang, S.~Ren, and J.~Sun, ``Deep residual learning for image
  recognition,'' in \emph{IEEE Conference on Computer Vision and Pattern
  Recognition}, 2016, pp. 770--778.

\bibitem{JCTVC-F158}
T.~Davies, ``Resolution switching for coding efficiency and resilience,''
  JCTVC-F158, 2011.

\bibitem{agbinya1992interpolation}
J.~Agbinya, ``Interpolation using the discrete cosine transform,''
  \emph{Electronics Letters}, vol.~28, no.~20, pp. 1927--1928, 1992.

\bibitem{serre2007robust}
T.~Serre, L.~Wolf, S.~Bileschi, M.~Riesenhuber, and T.~Poggio, ``Robust object
  recognition with cortex-like mechanisms,'' \emph{IEEE Transactions on Pattern
  Analysis and Machine Intelligence}, vol.~29, no.~3, pp. 411--426, 2007.

\bibitem{szegedy2015going}
C.~Szegedy, W.~Liu, Y.~Jia, P.~Sermanet, S.~Reed, D.~Anguelov, D.~Erhan,
  V.~Vanhoucke, and A.~Rabinovich, ``Going deeper with convolutions,'' in
  \emph{IEEE Conference on Computer Vision and Pattern Recognition}, 2015, pp.
  1--9.

\bibitem{dosovitskiy2015learning}
A.~Dosovitskiy, J.~Tobias~Springenberg, and T.~Brox, ``Learning to generate
  chairs with convolutional neural networks,'' in \emph{IEEE Conference on
  Computer Vision and Pattern Recognition}, 2015, pp. 1538--1546.

\bibitem{noh2015learning}
H.~Noh, S.~Hong, and B.~Han, ``Learning deconvolution network for semantic
  segmentation,'' in \emph{IEEE International Conference on Computer Vision},
  2015, pp. 1520--1528.

\bibitem{sun2003image}
J.~Sun, N.-N. Zheng, H.~Tao, and H.-Y. Shum, ``Image hallucination with primal
  sketch priors,'' in \emph{IEEE Conference on Computer Vision and Pattern
  Recognition}, vol.~2.\hskip 1em plus 0.5em minus 0.4em\relax IEEE, 2003, pp.
  II--729.

\bibitem{lee2009intra}
S.~H. Lee and N.~I. Cho, ``Intra prediction method based on the linear
  relationship between the channels for {YUV} 420 intra coding,'' in \emph{IEEE
  International Conference on Image Processing}.\hskip 1em plus 0.5em minus
  0.4em\relax IEEE, 2009, pp. 1037--1040.

\bibitem{zhang2014chroma}
X.~Zhang, C.~Gisquet, E.~Francois, F.~Zou, and O.~C. Au, ``Chroma intra
  prediction based on inter-channel correlation for {HEVC},'' \emph{IEEE
  Transactions on Image Processing}, vol.~23, no.~1, pp. 274--286, 2014.

\bibitem{li2016lambda}
L.~Li, B.~Li, D.~Liu, and H.~Li, ``$\lambda$-domain rate control algorithm for
  {HEVC} scalable extension,'' \emph{IEEE Transactions on Multimedia}, vol.~18,
  no.~10, pp. 2023--2039, 2016.

\bibitem{jia2014caffe}
Y.~Jia, E.~Shelhamer, J.~Donahue, S.~Karayev, J.~Long, R.~Girshick,
  S.~Guadarrama, and T.~Darrell, ``Caffe: Convolutional architecture for fast
  feature embedding,'' in \emph{ACM International Conference on
  Multimedia}.\hskip 1em plus 0.5em minus 0.4em\relax ACM, 2014, pp. 675--678.

\bibitem{bossen2011common}
F.~Bossen, ``Common test conditions and software reference configurations,''
  JCTVC-F900, 2011.

\bibitem{bjontegaard2001calcuation}
G.~Bjontegaard, ``Calcuation of average {PSNR} differences between
  {RD}-curves,'' VCEG-M33, 2001.

\bibitem{song2013sjtu}
L.~Song, X.~Tang, W.~Zhang, X.~Yang, and P.~Xia, ``The {SJTU 4K} video sequence
  dataset,'' in \emph{International Workshop on Quality of Multimedia
  Experience}.\hskip 1em plus 0.5em minus 0.4em\relax IEEE, 2013, pp. 34--35.

\end{thebibliography}

\end{document}